\pdfoutput=1
\documentclass[preprint,aps,prd,nofootinbib]{revtex4-2}
\usepackage{amssymb,amsmath,graphicx,color,microtype,enumerate,slashed,hyperref,xcolor,cases,verbatim}
\usepackage{tikz}
\usetikzlibrary{decorations.pathmorphing}
\tikzset{snake it/.style={decorate, decoration={snake}}}
\hypersetup{linktocpage=true,colorlinks=true,linkcolor=blue,filecolor=blue,urlcolor=blue, citecolor=blue}
\graphicspath{{fig/}}

\makeatletter
\def\@hangfrom@section#1#2#3{\@hangfrom{#1#2}#3}
\def\@hangfroms@section#1#2{#1#2}
\makeatother

\makeatletter
{\vskip200\p@\@plus1fil
	\penalty-200\relax
	\vskip-200\p@\@plus-1fil}
{}
{}
{}
\makeatother

\begin{document}
\count\footins = 1000 
\title{On the Decoherence of Primordial Gravitons}
\author{Sirui Ning$^{1}$}
\email{sirui.ning@physics.ox.ac.uk}
\author{Chon Man Sou$^{2,3}$}
\email{cmsou@connect.ust.hk}
\author{Yi Wang$^{2,3}$}
\email{phyw@ust.hk}
\affiliation{${}^1$The Rudolf Peierls Centre for Theoretical Physics, University of Oxford, Oxford OX1 3PU, United Kingdom}
\affiliation{${}^2$Department of Physics, The Hong Kong University of Science and Technology, \\
	Clear Water Bay, Kowloon, Hong Kong, P.R.China}
\affiliation{${}^3$The HKUST Jockey Club Institute for Advanced Study, The Hong Kong University of Science and Technology, \\
	Clear Water Bay, Kowloon, Hong Kong, P.R.China}

\begin{abstract}
It is well-known that the primordial scalar curvature and tensor perturbations, $\zeta$ and $\gamma_{ij}$, are conserved on super-horizon scales in minimal inflation models. However, their wave functional has a rapidly oscillating phase which is slow-roll unsuppressed, as can be seen either from boundary (total-derivative) terms of cosmological perturbations, or the WKB approximation of the Wheeler-DeWitt equation. Such an oscillatory phase involves gravitational non-linearity between scalar and tensor perturbations. By tracing out unobserved modes, the oscillatory phase causes faster decoherence of primordial gravitons compared to those by bulk interactions. Our results put a stronger lower bound of decoherence effect to the recent proposals probing squeezed primordial gravitons.
\end{abstract}

\maketitle
\pagebreak
\tableofcontents

\section{Introduction}
During inflation, the vacuum of the graviton evolves into a quantum squeezed state \cite{Grishchuk:1989ss,Grishchuk:1990bj,Albrecht:1992kf,Polarski:1995jg}, opening up possibilities to probe the non-classicality of gravitons via observations. There have been proposals to test the non-classicality of squeezed gravitons, including the quantum noise in gravitational-wave detectors and geodesics \cite{Parikh:2020nrd,Parikh:2020fhy,Parikh:2020kfh,Kanno:2020usf,Haba:2020jqs,Cho:2021gvg,Bak:2022oyn,Cho:2023dmh},\footnote{See also the related framework developed earlier in the semi-classical stochastic gravity \cite{Calzetta:1993qe,Hu:1999mm,Hu:2008rga} and the bound of squeezing from the current LIGO-Virgo data \cite{Hertzberg:2021rbl}.} indirect detection through decoherence of matter \cite{Kanno:2020usf,Kanno:2021gpt} and electromagnetic field \cite{Arani:2021oim}, the Hanbury Brown-Twiss interferometry with the related sub-Poissonian statistics \cite{Giovannini:2010xg,Kanno:2018cuk,Kanno:2019gqw} and the interaction with an optical cavity \cite{Guerreiro:2019vbq,Guerreiro:2021qgk}. In particular, the quantum noise produced by the primordial gravitational wave can be largely enhanced by the inflationary squeezed states, providing a chance to make the non-classicality of gravitons detectable. However, it is well-known that the cosmological perturbations, including scalar and tensor, can experience the quantum-to-classical transition through the interaction with environment during inflation, described by the environment-induced decoherence. As a first step to analyze the potential obstruction of the mentioned proposals, we study the decoherence of the primordial gravitons during the simplest single-field inflation.

The environment-induced decoherence of scalar perturbations has been widely studied with various interactions and tools \cite{Hu:1992ig,Calzetta:1995ys,Lombardo:2005iz,Kiefer:2006je,Kiefer:2008ku,Burgess:2006jn,Martineau:2006ki,Prokopec:2006fc,Burgess:2014eoa,Boyanovsky:2015tba,Nelson:2016kjm,Shandera:2017qkg,Martin:2018zbe,Martin:2018lin,DaddiHammou:2022itk,Colas:2022hlq,Ye:2018kty,Sou:2022nsd,Friedrich:2019hev,Brahma:2020zpk}, and some of the frameworks are also applicable to the decoherence of tensor perturbations. For the primordial gravitons, the possible sources of decoherence include the non-linear interaction between tensor modes \cite{Gong:2019yyz}, the scalar-tensor interaction \cite{Burgess:2022nwu,DaddiHammou:2022itk}, and the primordial magnetic field \cite{Kanno:2022ykw,Kanno:2022kve}. Focusing on the simplest single-field inflation, the decoherence effects of gravitons by scalar and tensor environments through the bulk interactions are comparable, and the quantum-to-classical transition happens at 7-9 e-folds after crossing the Hubble horizon,\footnote{We also compare with the results in \cite{Gong:2019yyz,Burgess:2022nwu}, and the related decoherence quantities are estimated in Sec. \ref{sec:compare_deco}.} proving the upper bound of the purity of the primordial squeezed gravitons.

As noted recently in \cite{Sou:2022nsd}, the slow-roll unsuppressed boundary (total-derivative) term 
\begin{align}
	\mathcal{L}_{{\rm bd},\zeta}=M_p^2\frac{d}{dt}\left(-2a^3He^{3\zeta}\right) \ , \label{eq:Lbd_zeta_previous}
\end{align}
which dominates for the super-horizon scalar curvature perturbation $\zeta$, can introduce rapidly oscillating non-Gaussian phase to the wave functional of cosmological perturbations, leading to much larger decoherence than the bulk interactions. Such boundary terms without any time derivative $\dot{\zeta}$ (similarly for tensor perturbation $\dot{\gamma}_{ij}$) cannot contribute to correlators $\langle \zeta^m \gamma^n\rangle$, so they are often neglected in the literature, whereas they also cannot be removed by field redefinitions \cite{Arroja:2011yj,Burrage:2011hd,Prokopec:2013zya}. In this paper, we extend the boundary-term decoherence to the case with tensor perturbation, which relies on the slow-roll unsuppressed scalar-tensor cubic boundary terms
\begin{align}
	\mathcal{L}_{{\rm bd},\zeta-\gamma}&=M_p^2\frac{d}{dt}\left[- \frac{a \partial_i\zeta  \partial_j\zeta  \gamma {}_i{}_j}{H}- \frac{a \zeta  \left(\partial_l\gamma {}_i{}_j \right)^2}{8 H}\right] \label{eq:scalar_tensor_boundary} \ , 
\end{align}
and we will show that these boundary terms also lead to larger decoherence than the bulk terms. 

There is a constant interest in discussing the Wheeler-DeWitt (WDW) equation \cite{DeWitt:1967yk,Wheeler:1968iap} and quantum gravity \cite{Witten:2022xxp,Araujo-Regado:2022gvw,McGough:2016lol,Hartman:2018tkw,Araujo-Regado:2022jpj}, which motivates us to discuss the relationship between these boundary terms and the wave functional obtained by the WDW equation. In the large volume limit of inflation \cite{Larsen:2004kf,Pimentel:2013gza,Pajer:2016ieg} and dS space \cite{Larsen:2003pf,Maldacena:2019cbz,Chakraborty:2023yed,Chakraborty:2023los} or similarly the asymptotic infinity in AdS \cite{Freidel:2008sh,Papadimitriou:2010as,Araujo-Regado:2022gvw}, the wave functional shares the same form, consisting of a real local action $W(h_{ij},\phi)$ and a part $Z(h_{ij},\phi)$ including non-local terms
\begin{align}
	\Psi(h_{ij},\phi)&=e^{iW(h_{ij},\phi)}Z(h_{ij},\phi) \ , \label{eq:wavefunction_W_Z}
\end{align}
where $h_{ij}$ is the spatial metric induced on a hypersurface, and $\phi$ is the inflaton in our case. With the form of the wave functional (\ref{eq:wavefunction_W_Z}), it is clear that the rapidly oscillating phase $W(h_{ij},\phi)$ cannot contribute to the expectation value of any observable defined by the spatial metric
\begin{align}
	\langle O(h_{ij})\rangle&=\int Dh_{ij} \left|\Psi(h_{ij},\phi)\right|^2 O(h_{ij}) \nonumber \\
	&=\int Dh_{ij} \left|Z(h_{ij},\phi)\right|^2 O(h_{ij}) \ ,
\end{align}
where we assume that $\Psi(h_{ij},\phi)$ is normalized and defined on the hypersurface with $\delta \phi=0$ (the $\zeta$-gauge), so all the quantum degrees of freedom are in the metric. The local action (or WKB phase) $W(h_{ij},\phi)$ can be calculated by applying the WKB approximation to the WDW equation \cite{Larsen:2004kf,Pimentel:2013gza,Pajer:2016ieg}, and we will show that it matches the non-Gaussian phase obtained from the boundary terms in the action of cosmological perturbations, thus contributing to the decoherence. It is noteworthy that the slow-roll suppressed scalar bulk interaction $\epsilon(\epsilon+\eta)a\left(\partial_i\zeta\right)^2\zeta$ studied in \cite{Nelson:2016kjm} also contributes a rapidly oscillating non-Gaussian phase at late time, thus the state is considered as the WKB type by the author, causing decoherence of $\zeta$ by tracing out unobserved modes. However, we will show that the WKB phase in the WDW state includes slow-roll unsuppressed parts involving both scalar $\zeta$ and tensor perturbations $\gamma_{ij}$.

The paper is organized as follows. In Sec. \ref{sec:bulk_boundary_cubic}, we review the setup of the simplest single-field inflation with a brief discussion of the choice of tensor perturbation and re-derive the splitting of bulk and boundary cubic terms. In Sec. \ref{sec:oscillating_phase}, we first discuss the non-Gaussian phase obtained from the slow-roll unsuppressed boundary terms in both the interaction and Schr\"odinger pictures, and we then check that it is consistent with the local phase obtained by solving the WDW equation with the WKB approximation. In Sec. \ref{eq:decoherence}, we calculate the decoherence effect of primordial gravitons by tracing out the unobserved scalar or tensor environments, confirming that the decoherence happens faster than the one caused by the bulk interactions. In Sec. \ref{sec:1_loop_correction}, we comment the one-loop quantum correction to the wave functional, including the contribution from the Faddeev-Popov ghost by diffeomorphism, and we argue that the correction is negligible in the leading decoherence effect. Sec. \ref{sec:conclusion} is the conclusion.

We set some notations for convenience. We label the comoving momenta of system modes with $\bf q$ and environment modes with $\bf k$, and the comoving momentum $\bf p$ can be used to label arbitrary modes. The integral modes with momentum conservation is denoted by $\int_{{\bf {p}}_1,\cdots,{\bf {p}}_n}=\int\frac{d^3p_1}{(2\pi)^3}\cdots\frac{d^3p_n}{(2\pi)^3}(2\pi)^3\delta^3({\bf p}_1+\cdots+{\bf p}_n)$, and integrating over two environment modes with a fixed system mode is $\int_{\bf {k}+\bf{k}'=-\bf{q}}=\int\frac{d^3k}{(2\pi)^3}\frac{d^3 k'}{(2\pi)^3}(2\pi)^3\delta^3(\bf {k}+\bf{k}'+\bf{q})$. We use a simpler notation $\sum_{s_1,\cdots,s_n}=\sum_{s_1,\cdots,s_n=+,-}$ to represent the sum of $n$ circular polarization indices.

\section{The bulk and boundary terms in the cubic order} \label{sec:bulk_boundary_cubic}
In this section, we review the setup for deriving the action of cosmological perturbations up to the cubic order \cite{Maldacena:2002vr} with a brief comment on the two common choices of tensor perturbation in the literature. Since the neglected temporal boundary terms obtained by integration by parts are not clearly shown in \cite{Maldacena:2002vr}, we also derive the splitting of bulk and boundary terms for both scalar and tensor perturbations.
\subsection{The setup} \label{sec:setup}
We start with the ADM decomposition of the metric
\begin{align}
	ds^2=-N^2dt^2+h_{ij}\left(dx^i+N^i dt\right)\left(dx^j+N^j dt\right) \ ,
\end{align}
where $N$ and $N^i$ are the lapse and shift respectively, and $h_{ij}$ is the spatial metric on the hypersurface which has the extrinsic curvature 
\begin{align}
	K_{\mu\nu}=\left(\delta^\rho_\mu+n^\rho n_\mu\right)\nabla_\rho n_\nu  \ ,
\end{align}
where $n^\mu$ is the normal of the hypersurface. We consider the simplest single-field inflation with the action
\begin{align}
	S&=\int_{\mathcal{M}}d^4x \sqrt{-g}\left[\frac{M_p^2}{2}R-\frac{1}{2}g^{\mu\nu}\partial_\mu\phi\partial_\nu\phi-V(\phi)\right]-\int_{\partial \mathcal{M}}d^3x M_p^2\sqrt{h}K \nonumber \\
	&=\frac{M_p^2}{2}\int_\mathcal{M}d^4x \sqrt{h}\left[N\left({^{(3)}}R-K^2+K_{ij}K^{ij}\right)+\frac{1}{N}\left(\dot{\phi}-N^i\partial_i\phi\right)^2-Nh^{ij}\partial_i\phi\partial_j\phi-2NV(\phi)\right] \ . \label{eq:action_single_field}
\end{align}
where the second term in the first line is the Gibbons-Hawking-York (GHY) boundary term for the manifold $\mathcal{M}$ \cite{York:1972sj,Gibbons:1976ue,York:1986lje} which cancels with the covariant derivative term when we decompose the 4-dimensional Ricci scalar
\begin{align}
	R&= {^{(3)}R}-K^2+K^\mu_\nu K^\nu_\mu-2\nabla_\mu\left(-K n^\mu+n^\nu \nabla_\nu n^\mu\right) \ ,
\end{align}
with ${^{(3)}R}$ the three-dimensional Ricci scalar. The Hubble parameter is determined by the uniform background of inflaton field as
\begin{align}
	3M_p^2H^2&=\frac{1}{2}\dot{\phi}^2+V(\phi) \ ,
\end{align}
with which the slow-roll parameters are defined
\begin{align}
	\epsilon=-\frac{\dot{H}}{H^2} \ , \ \eta=\frac{\dot{\epsilon}}{H\epsilon} \ ,
\end{align}
satisfying the slow-roll conditions $\epsilon\ll 1$ and $|\eta|\ll 1$.

In the literature, there are two definitions of tensor perturbations in the spatial metric $h_{ij}$ which are applied in the decoherence problems \cite{Gong:2019yyz,Burgess:2022nwu}, and here we briefly discuss the difference. One is $\gamma_{ij}$ defined in the comoving gauge ($\zeta$-gauge with $\delta\phi=0$), e.g. in \cite{Maldacena:2002vr}
\begin{align}
	h_{ij}&=a^2e^{2\zeta}(e^\gamma)_{ij} \ , \label{eq:spatial_metric_gamma}
\end{align}
where $\zeta$ is the scalar curvature perturbation, and $\gamma_{ij}$ satisfies $\gamma_{ii}=\partial_i{\gamma_{ij}}=0$. Another is defined by the scalar-vector-tensor decomposition \cite{Prokopec:2010be}
\begin{align}
	h_{ij}&=a^2\left(\delta_{ij}+\frac{h_S}{3}\delta_{ij}+h^{ TT}_{ij}\right) \ , \label{eq:spatial_metric}
\end{align}
which includes the scalar part $h_S$ and the transverse traceless part $h^{ TT}_{ij}$. The curvature perturbation $\zeta$ is related to $(h_S,h^{ TT}_{ij})$ on the hypersurface with $\delta\phi=0$ by comparing the same $\det(h_{ij})$ calculated in (\ref{eq:spatial_metric_gamma}) and (\ref{eq:spatial_metric})
\begin{align}
	\zeta\left(h_S,h^{ TT}_{ij}\right) &\approx \frac{h_S}{6}-\frac{1}{36}\left(h_S^2+3h^{TT}_{ij}h^{TT}_{ij}\right)+\mathcal{O}\left(h_S^3,\left(h_{ij}^{TT}\right)^3\right) \ ,
\end{align}
showing the dependence between $\zeta$ and $h_{ij}^{ TT}$, so $h_{ij}^{ TT}$ may not be a good choice when $\zeta$ is considered as the environment of decoherence of gravitons. Therefore, under the condition $\delta\phi=0$, one may either choose $(\zeta,\gamma_{ij})$ or $(h_S,h^{TT}_{ij})$ to be the scalar and tensor perturbations to make the calculation convenient and avoid some ambiguities.

On the other hand, $\gamma_{ij}$ defined in the first case conserves outside the horizon, as the cubic interactions related to $\gamma_{ij}$ all include derivatives which will be shown explicitly in (\ref{eq:2zeta1gamma_bulk})-(\ref{eq:3gamma_bulk}). However, $h^{ TT}_{ij}$ does not conserve outside the horizon as the cubic interactions include a term without any derivative \cite{Gong:2019yyz}
\begin{align}
	\mathcal{L}_{h^{TT}h^{TT}h^{TT}}\supset 2\left(1-\frac{\epsilon}{3}\right)M_p^2H^2a^3 h^{ TT}_{ij}h^{ TT}_{jl}h^{ TT}_{li} \ ,
\end{align}
suggesting that $\gamma_{ij}$ is more convenient to the describe the super-horizon evolution of tensor mode. As also discussed in \cite{Ota:2022xni}, gravitational wave should not perturb the spatial volume, and the property $\det (e^{\gamma})_{ij}=1$ also suggests that $\gamma_{ij}$ is more appropriate. 

Due to the mentioned reasons, we adopt the spatial metric defined with $\left(\zeta,\gamma_{ij}\right)$ (\ref{eq:spatial_metric_gamma}), consistent with the choice in \cite{Nelson:2016kjm,Sou:2022nsd,Burgess:2022nwu}. Varying  the action (\ref{eq:action_single_field}) with respect to $N$ and $N^i$ gives the constraint equations, solving them gives 
\begin{align}
	N=1+\frac{\dot{\zeta}}{H} \ , \	N^i=\frac{1}{a^2}\left(-\frac{\partial_i\zeta}{H}+\partial_i \chi\right) \ , \ \partial^2\chi=a^2\epsilon \dot{\zeta} \ .
\end{align}
With these, expanding the action (\ref{eq:action_single_field}) to the second order gives the free actions of scalar
\begin{align}
	S^{(\zeta)}_{2}&=M_p^2\int dt d^3x \ \epsilon a^3\left[\dot{\zeta}^2-a^{-2}\left(\partial_i\zeta\right)^2\right] \nonumber \\
	&=M_p^2\int dt \frac{d^3p}{(2\pi)^3} \ \epsilon a^3 \left(\dot{\zeta}_{\bf p}\dot{\zeta}_{-{\bf p}}-\frac{p^2}{a^2}\zeta_{\bf p}\zeta_{-{\bf p}}\right) \ , \label{eq:free_zeta_action}
\end{align}
and tensor perturbations
\begin{align}
	S^{(\gamma)}_{2}&=\frac{M_p^2}{8}\int dt d^3x \ a^3 \left[ \dot{\gamma}_{ij}\dot{\gamma}_{ij}-a^{-2} \left(\partial_l\gamma_{ij} \right)^2\right] \nonumber \\
	&=\frac{M_p^2}{2}\sum_{s}\int dt \frac{d^3p}{(2\pi)^3} \ a^3 \left(\dot{\gamma}^s_{\bf p}\dot{\gamma}^s_{-{\bf p}}-\frac{p^2}{a^2}\gamma^s_{\bf p}\gamma^s_{-{\bf p}}\right) \ , \label{eq:free_gamma_action}
\end{align}
where in the second line the following mode decomposition is applied \cite{Maldacena:2011nz}
\begin{align}
	\gamma_{ij}({\bf x},t) =\sum_{s}\int \frac{d^3p}{(2\pi)^3} \gamma^s_{\bf p}(t)e^s_{ij}({\bf p})e^{i{\bf p}\cdot {\bf x}} \ , 
\end{align}
and the symmetric polarization tensors satisfy 
\begin{align}
	e^{\pm}_{ij}(-{\bf p})&=\left(e^{\pm}_{ij}({\bf p})\right)^* \nonumber \\ 
	e^s_{ij}({\bf p})(e^{s'}_{ij}({\bf p}))^*&=4\delta_{s,s'} \nonumber \\
	e^{\pm}_{ii}({\bf p})&=0 \nonumber \\
	p_ie^{\pm}_{ij}({\bf p})&=0 \ .
\end{align}

By quantizing $\zeta$ and $\gamma_{ij}$, the free actions (\ref{eq:free_zeta_action}) and (\ref{eq:free_gamma_action}) imply a time-dependent Gaussian wave functional \cite{Maldacena:2002vr}
\begin{align}
	\Psi_G(\zeta,\gamma)&=N_G(\tau)\exp\left(-\frac{1}{2}\int \frac{d^3p}{(2\pi)^3}A^{(\zeta)}_p(\tau)\zeta_{\bf p}\zeta_{-{\bf p}}-\frac{1}{2}\sum_s\int \frac{d^3p}{(2\pi)^3}A^{(\gamma)}_p(\tau)\gamma^s_{\bf p}\gamma^s_{-{\bf p}}\right) \ , \label{eq:Gaussian_wave_fun}
\end{align}
where the normalization factor $N_G=\prod_{\bf p}N_{G,{\bf p}}^{(\zeta)}N_{G,{\bf p}}^{(\gamma)}$ is separable, and the coefficients\footnote{The time derivative of $\epsilon$ is neglected when we calculate $A^{(\zeta)}_p(\tau)$.}
\begin{align}
	A^{(\zeta)}_p(\tau)&=-2\epsilon M_p^2\frac{i}{(H\tau)^2}\frac{{u'}^{(\zeta)}_p(\tau)}{u^{(\zeta)}_p(\tau)} \nonumber 
	\\
	&=2p^3\frac{\epsilon M_p^2}{H^2}\frac{1-\frac{i}{p\tau}}{1+p^2\tau^2} \ , \label{eq:A_exponent_zeta}\\
	A^{(\gamma)}_p(\tau)&=-M_p^2\frac{i}{(H\tau)^2}\frac{{u'}^{(\gamma)}_p(\tau)}{{u}^{(\gamma)}_p(\tau)} \nonumber \\
	&=p^3\frac{M_p^2}{H^2}\frac{1-\frac{i}{p\tau}}{1+p^2\tau^2} \label{eq:A_exponent_gamma} \ ,
\end{align}
where $\tau=\int^t \frac{dt'}{a(t')}$ is the conformal time with ${}'=\frac{d}{d\tau}$, and the mode functions are 
\begin{align}
	u^{(\zeta)}_p(\tau)&=\frac{H}{2M_p\sqrt{\epsilon p^3}}(1-ip\tau)e^{ip\tau}\nonumber \\
	u^{(\gamma)}_p(\tau)&= \frac{H}{M_p\sqrt{2p^3}}(1-ip\tau)e^{ip\tau} . \label{eq:mode_function}
\end{align}
For the Gaussian wave functional (\ref{eq:Gaussian_wave_fun}), the scalar and tensor power spectra are related to the two coefficients (\ref{eq:A_exponent_zeta}) and (\ref{eq:A_exponent_gamma}) as
\begin{align}
	P^{(\zeta,\gamma)}_p(\tau)&=\frac{1}{2{\rm Re}A^{(\zeta,\gamma)}_p(\tau)} \ .
\end{align}
At the quadratic level, all the modes evolve independently, and thus we need the cubic order action for the interactions between observed and unobserved modes.

\subsection{Splitting the bulk and boundary cubic terms}
We want to find out all the scalar and tensor temporal boundary terms neglected in \cite{Maldacena:2002vr}, as they will be shown to be important to the inflationary decoherence. The guideline of splitting the bulk and boundary terms in the $\zeta$-gauge is to make the former matches the one derived in the $\delta\phi$-gauge (spatially flat) (see also \cite{Rigopoulos:2011eq,Prokopec:2012ug,Prokopec:2013zya} by arranging the bulk terms in the gauge-invariant manner), and it is expected that all the cubic bulk terms involving $\zeta$ is slow-roll suppressed since the transformation $\delta\phi\approx-\frac{\dot{\phi}}{H}\zeta\sim \mathcal{O}(\sqrt{\epsilon})\zeta$ introduces slow-roll parameters. On the other hand, since the tensor perturbations in the two gauges are on the same slow-roll order, bulk terms include some slow-roll unsuppressed $\gamma\gamma\gamma$ interactions.

The splitting of the bulk and boundary interactions can be found in the literature \cite{Burrage:2011hd,Garcia-Saenz:2019njm,Prokopec:2013zya}, and here we derive again with the notations used in this paper. With the Mathematica package {\it MathGR} \cite{Wang:2013mea} doing integration by parts, we obtain the following splitting of bulk and boundary cubic terms from the action (\ref{eq:action_single_field}) (with spatial total derivative neglected)
\begin{align}
	\mathcal{L}^{(3)}=\mathcal{L}_{\zeta\zeta\zeta}+\mathcal{L}_{\zeta\zeta\gamma}+\mathcal{L}_{\zeta\gamma\gamma}+\mathcal{L}_{\gamma\gamma\gamma}+f(\zeta,\gamma)\frac{\delta L_2}{\delta \zeta}+f_{ij}(\zeta,\gamma)\frac{\delta L_2}{\delta \gamma_{ij}}+\mathcal{L}_{\rm bd,\zeta\zeta\zeta}+\mathcal{L}_{\rm bd,\zeta\zeta\gamma}+\mathcal{L}_{\rm bd,\zeta\gamma\gamma} \ , \label{eq:splitting_bulk_boundary}
\end{align}
where the four bulk terms
\begingroup
\allowdisplaybreaks
\begin{align}
	\mathcal{L}_{\zeta\zeta\zeta}&=M_p^2\left[a^3\epsilon(\epsilon-\eta)\zeta\dot{\zeta}^2+a\epsilon(\epsilon+\eta)\zeta(\partial_i\zeta)^2+\left(\frac{\epsilon}{2}-2\right)\frac{\partial^2\chi}{a}\partial_i\chi\partial_i\zeta+\frac{\epsilon}{4a}\partial^2\zeta\left(\partial_i\chi\right)^2\right] \label{eq:3zeta_bulk} \\
	\mathcal{L}_{\zeta\zeta\gamma}&=M_p^2\left[-\frac{1}{2} a \epsilon  \chi  \partial_i\partial_j\zeta  \dot\gamma_i{}_j
	+ \frac{\partial_i\chi  \partial_j\chi  \partial^2\gamma_i{}_j}{4 a}
	+ a \epsilon  \partial_i\zeta  \partial_j\zeta  \gamma_i{}_j\right] \label{eq:2zeta1gamma_bulk} \\
	\mathcal{L}_{\zeta\gamma\gamma}&=M_p^2\left[\frac{1}{8} a^3 \epsilon  \zeta  \dot\gamma {}_i{}_j {}^2
	- \frac{1}{4} a \partial_l\chi  \dot\gamma {}_i{}_j \partial_l\gamma{}_i{}_j
	+ \frac{1}{8} a \epsilon  \zeta  \left(\partial_l\gamma {}_i{}_j \right)^2\right] \label{eq:1zeta2gamma_bulk}\\
	\mathcal{L}_{\gamma\gamma\gamma}&=M_p^2\left[\frac{1}{4} a \partial_m\gamma {}_i{}_l \partial_l\gamma {}_j{}_m 
	\gamma {}_i{}_j
	+ \frac{1}{8} a \partial_i\gamma {}_l{}_m \partial_j\gamma {}_l{}_m 
	\gamma {}_i{}_j\right] \label{eq:3gamma_bulk} \ ,
\end{align}
\endgroup
the two EOM terms\footnote{Note that the last term of (\ref{eq:EOM_term_f}) involves $\partial^{-2}$, which agrees with (3.22) in \cite{Prokopec:2013zya}.}
\begingroup
\allowdisplaybreaks
\begin{align}
	f(\zeta,\gamma)&=-\frac{\dot{\zeta}\zeta}{H}+\frac{1}{4a^2H^2}\left[\left(\partial_i\zeta\right)^2-\partial^{-2}\partial_i\partial_j\left(\partial_i\zeta\partial_j\zeta\right)\right]-\frac{1}{2a^2H}\left[\partial_i\zeta\partial_i\chi-\partial^{-2}\partial_i\partial_j\left(\partial_i\zeta\partial_j\chi\right)\right] \nonumber \\
	&+\frac{\partial_i\partial_j\zeta  \dot\gamma {}_i{}_j}{4 H}\partial^{-2} \label{eq:EOM_term_f} \\ 
	f_{ij}(\zeta,\gamma)&=- \frac{ \zeta  \dot\gamma {}_i{}_j}{H}+\frac{\partial_i\zeta  \partial_j\zeta }{a^2 H^2}
	+ \frac{2 \chi  \partial_i\partial_j\zeta }{a^2 H}\\
	\frac{\delta L_2}{\delta \zeta}&=2M_p^2\left[-\frac{d}{dt}\left(\epsilon a^3 \dot{\zeta}\right)+\epsilon a\partial^2\zeta\right] \\
		\frac{\delta L_2}{\delta \gamma_{ij}}&= \frac{M_p^2}{4}\left[-\frac{d}{dt}\left(a^3\dot{\gamma}_{ij}\right)+a\partial^2\gamma_{ij}\right]\ ,
\end{align}
\endgroup
and the boundary terms
\begingroup
\allowdisplaybreaks
\begin{align}
	\mathcal{L}_{\rm bd,\zeta\zeta\zeta}&=M_p^2\frac{d}{dt}\Bigg\{-9a^3H\zeta^3+\frac{a}{H}\left(1-\epsilon\right)\zeta\left(\partial_i\zeta\right)^2-\frac{1}{4aH^3}\left(\partial_i\zeta\right)^2\partial^2\zeta \nonumber \\
	&- \frac{\epsilon a^3}{H}\zeta\dot{\zeta}^2-\frac{\zeta}{2aH}\left[\left(\partial_i\partial_j\chi\right)^2-\left(\partial^2\chi\right)^2\right]+\frac{\zeta}{2aH^2}\left(\partial_i\partial_j\zeta\partial_i\partial_j\chi-\partial^2\zeta\partial^2\chi\right)\Bigg\} \label{eq:3zeta_bd}\\
	\mathcal{L}_{\rm bd,\zeta\zeta\gamma}&=M_p^2\frac{d}{dt}\left(- \frac{a \partial_i\zeta  \partial_j\zeta  \gamma {}_i{}_j}{H}+\frac{a \partial_i\zeta  \partial_j\zeta  \dot\gamma {}_i{}_j}{4 H^2}
	+ \frac{a \chi  \partial_i\partial_j\zeta  \dot\gamma {}_i{}_j}{2 H}
	\right) \label{eq:2zeta1gamma_bd} \\
	\mathcal{L}_{\rm bd,\zeta\gamma\gamma}&=M_p^2\frac{d}{dt} \left[- \frac{a \zeta  \left(\partial_l\gamma {}_i{}_j \right)^2}{8 H}-\frac{a^3 \zeta  \dot\gamma {}_i{}_j {}^2}{8 H} \right] \label{eq:1zeta2gamma_bd} \ .
\end{align}
\endgroup

Using the size of power spectra of scalar and tensor perturbations
\begin{align}
	\Delta^2_\zeta&=\lim_{\tau\to 0}\frac{p^3}{2\pi^2}P^{(\zeta)}_p(\tau)=\frac{H^2}{8\pi^2\epsilon M_p^2}  \ , \\ \Delta^2_\gamma&=\lim_{\tau\to 0}\frac{p^3}{2\pi^2}\sum_{s,s'}P^{(\gamma)}_p(\tau)4\delta_{s,s'}=\frac{2H^2}{\pi^2 M_p^2} \ , \label{eq:Delta_square}
\end{align}
we expect that the tensor perturbation is slow-roll suppressed compared to the scalar perturbation $\gamma\sim \mathcal{O}(\sqrt\epsilon) \zeta$. Table \ref{tab:int_terms}\footnote{In the choice of $h^{ TT}_{ij}$, the leading three-tensor interaction is $a^3h_{ij}^{ TT}h_{jl}^{ TT}h_{li}^{ TT}$, but we have chosen $\gamma_{ij}$.} is the summary of the size of interaction terms which do not involve any time derivative,\footnote{Terms with time derivatives are usually neglected for decoherence. For bulk terms, time derivatives are mainly contributed by sub-horizon modes which cause sub-dominated decoherence \cite{Burgess:2022nwu}. For boundary terms with time derivatives, they can be removed by field redefinitions \cite{Burrage:2011hd}.} and only the most dominated terms of each type are shown. It is clear that boundary terms are less slow-roll suppressed compared to the bulk terms.
\begin{table}
	\begin{tabular}{|c |c |c |c| }
		\hline
		bulk/boundary & type & leading interaction of each type & order \\
		\hline
		bulk & $\zeta\zeta\zeta$ & $\epsilon(\epsilon+\eta) a (\partial_i \zeta)^2\zeta$ & $\epsilon(\epsilon+\eta) \zeta^3$ \\		
		\hline
		bulk & $\zeta\zeta\gamma$ & $\epsilon a \partial_i\zeta \partial_j\zeta\gamma_{ij}$ & $\epsilon^{\frac{3}{2}}\zeta^3$ \\
		\hline
		bulk & $\zeta\gamma\gamma$ &	$\epsilon a \zeta \partial_l\gamma_{ij}\partial_l\gamma_{ij}$ & $\epsilon^2 \zeta^3$ \\
		\hline
		bulk & $\gamma\gamma\gamma$ & $a \partial_i\gamma {}_l{}_m \partial_j\gamma {}_l{}_m \gamma {}_i{}_j $  & $\epsilon^{\frac{3}{2}}\zeta^3$  \\
		\hline
		boundary & $\zeta\zeta\zeta$ & $\partial_t(a^3 \zeta^3)$ & $\zeta^3$ \\
		\hline
		boundary & $\zeta\zeta\gamma$ & $\partial_t(a \partial_i \zeta \partial_j \zeta \gamma_{ij})$ & $\epsilon^{\frac{1}{2}}\zeta^3$ \\
		\hline
		boundary & $\zeta\gamma\gamma$ & $\partial_t(a\zeta \partial_l\gamma_{ij} \partial_l\gamma_{ij})$ & $\epsilon \zeta^3$ \\
		\hline
	\end{tabular}
	\caption{The leading interaction terms of each type and their orders of magnitudes. \label{tab:int_terms}}
\end{table}

\section{The oscillating phase of inflationary wave functional} \label{sec:oscillating_phase}
It has been shown in the literature \cite{Arroja:2011yj,Burrage:2011hd,Garcia-Saenz:2019njm} that total time derivative terms in the Lagrangian $\mathcal{L}_{\rm int} \supset -\partial_t \mathcal{K}$ generally contribute to the in-in correlators as\footnote{The relation $H_{\rm int}=-L_{\rm int}$ holds in the cubic order with the interaction picture \cite{Wang:2013zva}.}
\begin{align}
	\langle O(t) \rangle &=\left\langle 0\left|\bar{T}\exp\left(i\int d^3x\int^t_{t_i}dt' \ \partial_{t'}\mathcal{K}\left(t'\right)\right) O(t)T\exp\left(-i\int d^3x\int^t_{t_i}dt' \ \partial_{t'}\mathcal{K}\left(t'\right)\right) \right|0\right\rangle \nonumber \\
	&=\left\langle 0\left|\exp\left(i\int d^3x \ \mathcal{K}(t)\right) O(t) \exp\left(-i\int d^3x \ \mathcal{K}(t)\right)\right|0\right\rangle \nonumber \\
	&\approx i \int d^3x\langle 0|\left[\mathcal{K}(t),O(t)\right] |0\rangle +\mathcal{O}(\mathcal{K}^2) \ . \label{eq:In_In_boundary}
\end{align}
Since decoherence is mainly contributed by terms without time derivatives \cite{Burgess:2022nwu}, we focus on the boundary terms involving $\zeta$, $\gamma_{ij}$ and their spatial derivatives, denoted by $\mathcal{K}\left(\zeta,\gamma,t\right)$ with the explicit forms shown in (\ref{eq:3zeta_bd})-(\ref{eq:1zeta2gamma_bd}). $\mathcal{K}\left(\zeta,\gamma,t\right)$ commutes with normal correlators of the form $O(t)=\zeta^m(t)\gamma^n(t)$, and the expectation value in (\ref{eq:In_In_boundary}) vanishes, so such boundary terms are often neglected in the literature. In this section, we will show how such boundary terms contribute a non-Gaussian phase to the wavefunctional of cosmological perturbations with three methods. Sec. \ref{sec:interaction_pic_boundary} and \ref{sec:Schrodinger_pic_boundary} calculate the phase with the interaction and Schr\"{o}dinger pictures respectively, and the results will be shown to be consistent in (\ref{eq:wavefun_interaction_pic}) and (\ref{eq:wavefun_schrodinger_pic}). In Sec. \ref{sec:WKB_phase_WDW}, we show that such a phase also matches the WKB approximation of the WDW equation, indicated by the WKB phase $W(h_{ij},\phi)$ in (\ref{eq:wavefunction_W_Z}), and the explicit result will be shown in (\ref{eq:W3_cubic_HJ}).

\subsection{The interaction picture} \label{sec:interaction_pic_boundary}
It is straightforward to generalize the interaction picture approach in \cite{Sou:2022nsd} to the case with tensor perturbation $\gamma_{ij}$, and the evolution operator is
\begin{align}
	U(\tau,\tau_i)&=U_0(\tau,\tau_i)T\exp\left(-i\int d^3x\int^t_{t_i}dt' \ \partial_{t'}\mathcal{K}\left(\zeta_I,\gamma_I,t'\right)\right) \nonumber \\
	&=U_0(\tau,\tau_i)\sum_{n=0}^\infty \frac{(-i)^n}{n!} \left(\mathcal{K}(\zeta_I,\gamma_I,\tau)\right)^n \nonumber \\
	&=\sum_{n=0}^\infty \frac{(-i)^n}{n!} \left(U_0(\tau,\tau_i)\mathcal{K}(\zeta_I,\gamma_I,\tau)U^{-1}_0(\tau,\tau_i)\right)^n U_0(\tau,\tau_i)\nonumber \\
	&=\exp\left(-i\int d^3x \ \mathcal{K}\left(\zeta_S,\gamma_S,\tau\right)\right)U_0(\tau,\tau_i) \ , \label{eq:Unitrary_operator}
\end{align}
where $U_0(\tau,\tau_i)=U^{(\zeta)}_0(\tau,\tau_i)U^{(\gamma)}_0(\tau,\tau_i)$ is the free evolution operator, and the labels $I$ and $S$ are the interaction and Schr\"{o}dinger pictures respectively. Applying $U(\tau,\tau_i)$ (\ref{eq:Unitrary_operator}) to the initial Gaussian state $|\Psi_G(\tau_i)\rangle$ gives a phase to the Schr\"{o}dinger wave functional
\begin{align}
	\left\langle \zeta,\gamma|\Psi(\tau)\right\rangle=\exp\left(-i\int d^3x \ \mathcal{K}\left(\zeta,\gamma,\tau\right)\right)\Psi_G(\zeta,\gamma) \ . \label{eq:wavefun_interaction_pic}
\end{align}

\subsection{The Schr\"{o}dinger picture} \label{sec:Schrodinger_pic_boundary}
We can also canonically quantize the theory with the boundary terms\footnote{We thank Haipeng An's suggestion of discussing the boundary term with the Hamiltonian formalism when CM Sou gave a seminar talk of \cite{Sou:2022nsd} at Tsinghua University.}
\begin{align}
	\mathcal{L}=\mathcal{L}_2^{(\zeta)}+\mathcal{L}_2^{(\gamma)}-\partial_t \mathcal{K} \ ,
\end{align}
with 
\begin{align}
	\mathcal{L}_2^{(\zeta)}+\mathcal{L}_2^{(\gamma)}&=f_{aa}(t)\dot \alpha^a\dot\alpha^a+j_{aa}(t)\alpha^a \alpha^a  \ , \nonumber \\
	\mathcal{K}(\zeta,\gamma,t)&=F_{abc}(t)\alpha^a\alpha^b\alpha^c \ , \label{eq:alpha_symbols}
\end{align}
where we use a more compact notation $\vec{\alpha}=\alpha^a$ to denote the dynamic fields $\zeta$ and $\gamma_{ij}$ for making the derivation simpler and more generic, so $j_{aa}$ and $F_{abc}$ can include comoving momenta and polarization tensors expressed in the Fourier space. With these notations, the total time derivative has the form
\begin{align}
	\partial_t \mathcal{K}=\dot F_{abc}\alpha^a\alpha^b\alpha^c+F_{abc}\left(\dot\alpha^a\alpha^b\alpha^c+\alpha^a\dot\alpha^b\alpha^c+\alpha^a\alpha^b\dot\alpha^c\right) \ ,
\end{align}
and the conjugate momenta are defined as
\begin{align}
	\Pi_a&=\frac{\partial\mathcal{L}}{\partial\dot{\alpha}^a} \nonumber \\
	&=2f_{bb}\delta^b_a\dot\alpha^b-\left(F_{dbc}+F_{bdc}+F_{bcd}\right)\delta^d_a\alpha^b\alpha^c\nonumber \\
	&=-i\frac{\delta}{\delta \alpha^a} \ ,
\end{align}
leading to the Hamiltonian density
\begin{align}
	\mathcal{H}&=\dot\alpha^a\Pi_a-\mathcal{L} \nonumber \\
	&=f_{aa}\dot \alpha^a\dot\alpha^a-j_{aa}\alpha^a \alpha^a+\dot F_{abc}\alpha^a\alpha^b\alpha^c \nonumber \\
	&=\frac{1}{4f_{aa}}\left[-i\frac{\delta}{\delta \alpha^a}+\bar{F}_{abc}\alpha^b\alpha^c\right]\left[-i\frac{\delta}{\delta \alpha^a}+\bar{F}_{abc}\alpha^b\alpha^c\right]-j_{aa}\alpha^a \alpha^a+\dot F_{abc}\alpha^a\alpha^b\alpha^c \ ,
\end{align}
where $\bar{F}_{abc}=F_{abc}+F_{bac}+F_{bca}$. We can easily check that the wavefunctional with the non-Gaussian phase (\ref{eq:wavefun_interaction_pic}), rewritten with the compact notations
\begin{align}
	\Psi({\vec \alpha})&=N(t) e^{-\int \frac{A_a}{2}\alpha^a\alpha^a}e^{-i\int F_{abc}\alpha^a\alpha^b\alpha^c} \ , \label{eq:wavefun_schrodinger_pic}
\end{align}
where $\int$ denotes integrals in the Fourier space, satisfies the Schr\"{o}dinger equation
\begin{align}
 	i\partial_t\Psi&=\left(i\dot{N}-iN\int \frac{\dot{A}_a}{2}\alpha^a\alpha^a+N\int \dot F_{abc}\alpha^a\alpha^b\alpha^c\right)e^{-\int \frac{A_a}{2}\alpha^a\alpha^a}e^{-i\int F_{abc}\alpha^a\alpha^b\alpha^c} \ , \nonumber \\
 	\int \mathcal{H}\Psi&=\left[\int \frac{A_a}{4f_{aa}}-\left(\frac{A^2_a}{4f_{aa}}+j_{aa}\right)\alpha^a\alpha^a+\int \dot F_{abc}\alpha^a\alpha^b\alpha^c\right] N e^{-\int \frac{A_a}{2}\alpha^a\alpha^a}e^{-i\int F_{abc}\alpha^a\alpha^b\alpha^c}\ , \nonumber \\
 	i\partial_t\Psi&=\int \mathcal{H}\Psi \ . \label{eq:Schrodinger_pic_match}
\end{align}
For the first two lines in (\ref{eq:Schrodinger_pic_match}), we applied the conditions obtained from the free theory (independent to $F_{abc}$) \cite{Burgess:2014eoa}
\begin{align}
	i\dot{N}=\int \frac{A_a}{4f_{aa}} N \ , \quad
	i\frac{\dot{A}_a}{2}=\frac{A^2_a}{4f_{aa}}+j_{aa} \ ,
\end{align}
and the following relations
\begin{align}
	\left[-i\frac{\delta}{\delta \alpha^a}+\bar{F}_{abc}\alpha^b\alpha^c\right]\Psi({\vec \alpha})&=	iA_a\alpha^a\Psi({\vec \alpha}) \ , \nonumber \\
	\left[-i\frac{\delta}{\delta \alpha^a}+\bar{F}_{abc}\alpha^b\alpha^c\right]^2\Psi({\vec \alpha})&=\left(A_a-A_a^2\alpha^a\alpha^a\right)\Psi({\vec \alpha}) \ .
\end{align}

\subsection{The WKB approximation of the Wheeler-DeWitt equation} \label{sec:WKB_phase_WDW}
The boundary terms can also be obtained by applying the WKB approximation to the wave function of universe $\Psi(h_{ij},\phi)$, obtained with the WDW equation \cite{DeWitt:1967yk,Wheeler:1968iap}. In \cite{Pimentel:2013gza,Pajer:2016ieg}, the WDW equation of gravity with a scalar field has been applied to analyze the consistency relation and bispectrum, and we follow the formalism. We start with the Hamiltonian corresponding to the action (\ref{eq:action_single_field}), defined on the hypersurface $\Sigma$ inducing $h_{ij}$,
\begin{align}
	H&=\int_\Sigma d^3x\Bigg\{ N\left[\frac{1}{2\kappa\sqrt{h}}G_{ij,kl}\pi^{ij}\pi^{kl}-\kappa\sqrt{h}\ {^{(3)}}R+\frac{1}{4\kappa\sqrt{h}}\pi_\phi^2+\kappa\sqrt{h}\left(h^{ij}\partial_i\phi\partial_j\phi+2V(\phi)\right)\right] \nonumber \\
	&+2\nabla_iN_j\pi^{ij}+h^{ij}N_j\partial_i\phi \pi_\phi \Bigg\} \ , \label{eq:Hamitlonian_single_field_inflation}
\end{align}
where $\kappa=\frac{M_p^2}{2}$, the conjugate momenta are
\begin{align}
	\pi^{ij}&=\frac{M_p^2}{2}\sqrt{h}\left(K^{ij}-h^{ij}K\right) \nonumber \\
	\pi_\phi&=M_p^2\frac{\sqrt{h}}{N}\left(\dot{\phi}-N^i\partial_i\phi\right) \ , \label{eq:conjugate_momenta_single_field}
\end{align}
and the DeWitt metric is
\begin{align}
	G_{ij,kl}=h_{ik}h_{jl}+h_{il}h_{jk}-h_{ij}h_{kl} \ .
\end{align}
By promoting the conjugate momenta (\ref{eq:conjugate_momenta_single_field}) to functional derivatives $\left(-i\hbar \frac{\delta}{\delta h_{ij}},-i\hbar \frac{\delta}{\delta \phi}\right)$ ($\hbar$ is restored for the moment) and varying (\ref{eq:Hamitlonian_single_field_inflation}) with respect to $\left(N,N_i\right)$, 
we obtain the Hamiltonian and momentum constraints for the wave functional $\Psi(h_{ij},\phi)$
\begin{align}
	\left[\frac{\hbar^2}{2\kappa\sqrt{h}}G_{ij,kl}\frac{\delta}{\delta h_{ij}}\frac{\delta}{\delta h_{kl}}+\kappa\sqrt{h}\ ^{(3)}R+\frac{\hbar^2}{4\kappa\sqrt{h}}\frac{\delta^2}{\delta \phi^2}-\kappa\sqrt{h}\left(h^{ij}\partial_i\phi\partial_j\phi+2V(\phi)\right)\right]\Psi&=0 \nonumber \\
	-2\hbar\nabla_i\left(\frac{1}{\sqrt{h}}\frac{\delta \Psi}{\delta h_{ij}}\right)+\frac{\hbar}{\sqrt{h}}h^{ij}\partial_i\phi\frac{\delta \Psi}{\delta\phi}&=0  \ , \label{eq:WdW_constraints}
\end{align}
where the first line is the WDW equation.

Here we adopt this formalism to analyze the dominated non-Gaussian phase of the wave functional, contributing to the decoherence. By applying the WKB approximation to the Hamiltonian constraint (\ref{eq:WdW_constraints}) with the ansatz $\Psi(h_{ij},\phi)=\exp\left(iW(h_{ij},\phi)/\hbar\right)$, the phase $W(h_{ij},\phi)$ satisfies the Hamilton-Jacobi (HJ) equation \cite{Larsen:2003pf,Larsen:2004kf} 
\begin{align}
	-\frac{1}{2\kappa\sqrt{h}}G_{ij,kl}\frac{\delta W}{\delta h_{ij}}\frac{\delta W}{\delta h_{kl}}+\kappa\sqrt{h}{}^{(3)}R-\frac{1}{4\kappa\sqrt{h}}\left(\frac{\delta W}{\delta \phi}\right)^2-\kappa\sqrt{h}\left(h^{ij}\partial_i\phi\partial_j\phi+2V(\phi)\right)&=0 \ , \label{eq:H_J_eq}
\end{align}
suggesting that the functional derivatives match the classical conjugate momenta (\ref{eq:conjugate_momenta_single_field}) \cite{Pajer:2016ieg}
\begin{align}
	\frac{\delta W}{\delta h_{ij}}&=\pi^{ij} \ , \quad	\frac{\delta W}{\delta \phi}=\pi_\phi \ .
\end{align}
The solution of the HJ equation (\ref{eq:H_J_eq}) has been constructed in the literature \cite{Salopek:1992qy,Parry:1993mw,Larsen:2003pf,Larsen:2004kf}, and its form up to terms with two spatial derivatives is 
\begin{align}
	W(h_{ij},\phi)&=M_p^2\int_\Sigma d^3x \sqrt{h}\left(U(\phi)+M(\phi)h^{ij}\partial_i\phi\partial_j\phi+\Phi(\phi){}^{(3)}R\right) \ ,
\end{align}
where
\begin{align}
	U(\phi)&=-\frac{2}{M_p}\sqrt{\frac{V(\phi)}{3-\epsilon_V}}= -2H \nonumber \\
	M(\phi)&=\frac{1}{M_p^2}\frac{1+2\eta_V-5\epsilon_V}{U(\phi)}\approx -\frac{1}{2HM_p^2}+\mathcal{O}(\epsilon,\eta)\nonumber \\
	\Phi(\phi)&=\frac{\epsilon_V-1}{U(\phi)} \approx \frac{1}{2H}+\mathcal{O}(\epsilon) \ ,
\end{align}
with $\epsilon_V=\frac{M_p^2}{2}\left(\frac{\partial_\phi V(\phi)}{V(\phi)}\right)^2$ and $\eta_V=M_p^2\frac{\partial^2_\phi V(\phi)}{V(\phi)}$ defined by the inflaton potential.
For $\Sigma$ defined in the $\zeta$-gauge with $\phi({\bf x},t)=\phi(t)$, the phase includes the cubic terms
\begin{align}
	&W_3(\zeta,\gamma_{ij}) =M_p^2\int_\Sigma d^3x\Bigg[-9 a^3 H \zeta ^3
	+ \frac{a \zeta  \left(\partial_i\zeta \right)^2}{H}
	- \frac{a \zeta  \left(\partial_l\gamma {}_i{}_j \right)^2}{8 H}
	- \frac{a \partial_i\zeta  \partial_j\zeta  \gamma {}_i{}_j}{H} \nonumber \\
	&+ \frac{a \partial_m\gamma {}_i{}_l \partial_l\gamma {}_j{}_m \gamma 
		{}_i{}_j}{4 H}
	+ \frac{a \partial_i\gamma {}_l{}_m \partial_j\gamma {}_l{}_m \gamma 
		{}_i{}_j}{8 H}+\mathcal{O}(\epsilon)\Bigg] \ , \label{eq:W3_cubic_HJ}
\end{align}
where the first line matches all the slow-roll unsuppressed boundary terms (\ref{eq:3zeta_bd})-(\ref{eq:1zeta2gamma_bd}) up to two spatial derivatives.\footnote{Note that the boundary terms with $\dot{\zeta}$ or $\dot{\gamma}_{ij}$ can be removed by field redefinitions \cite{Arroja:2011yj,Burrage:2011hd}, and thus we do not consider them.} The $\gamma\gamma\gamma$ terms in the second line has the same form with the bulk terms $\mathcal{L}_{\gamma\gamma\gamma}$ (\ref{eq:3gamma_bulk}) which come from $^{(3)}R$ \cite{Maldacena:2011nz} since the bulk terms mainly contribute a real phase to the wave functional at late time, and this can be shown as follows.

As a supplement to the first two methods, we consider the remaining slow-roll unsuppressed bulk term $\mathcal{L}_{\gamma\gamma\gamma}$ (\ref{eq:3gamma_bulk}), which contributes a non-Gaussian part to the Schr\"odinger wave functional as \cite{Maldacena:2002vr,Maldacena:2011nz,Nelson:2016kjm}
\begin{align}
	&\Psi_{NG}^{(\gamma\gamma\gamma)} \nonumber \\
	&\approx\exp\left[\sum_{s_1,s_2,s_3}\int_{{\bf p}_1,{\bf p}_2,{\bf p}_3} i \int^{\tau}_{\tau_i}\frac{d\tau'}{H\tau'}\tilde{H}^{(\gamma\gamma\gamma)}{}^{s_1,s_2,s_3}_{{\bf p}_1,{\bf p}_2,{\bf p}_3}(\tau')\frac{u^{(\gamma)}_{p_1}(\tau')u^{(\gamma)}_{p_2}(\tau')u^{(\gamma)}_{p_3}(\tau')}{u^{(\gamma)}_{p_1}(\tau)u^{(\gamma)}_{p_2}(\tau)u^{(\gamma)}_{p_3}(\tau)}\gamma^{s_1}_{{\bf p}_1}\gamma^{s_2}_{{\bf p}_2}\gamma^{s_3}_{{\bf p}_3}\right] \ , \label{eq:NG_3gamma}
\end{align}
where 
\begin{align}
	&\tilde{H}^{(\gamma\gamma\gamma)}{}^{s_1,s_2,s_3}_{{\bf p}_1,{\bf p}_2,{\bf p}_3}(\tau') \nonumber \\
	&=\frac{M_p^2}{8}a(\tau')\left[2p_{1,m}p_{2,l}e^{s_1}_{il}({\bf p}_1)e^{s_2}_{jm}({\bf p}_2)e^{s_3}_{ij}({\bf p}_3)+p_{1,i}p_{2,j}e^{s_1}_{lm}({\bf p}_1)e^{s_2}_{lm}({\bf p}_2)e^{s_3}_{ij}({\bf p}_3)\right] \ ,
\end{align}
and  $u_p^{(\gamma)}(\tau)$ is defined in (\ref{eq:mode_function}). The time integral in (\ref{eq:NG_3gamma}) is proportional to
\begin{align}
	&I^{(\gamma\gamma\gamma)}_{{\bf p}_1,{\bf p}_2,{\bf p}_3}(\tau) \nonumber \\
	&=-i\int^{\tau}_{\tau_i}\frac{d\tau'}{H^2\tau'^2}\frac{u^{(\gamma)}_{p_1}(\tau')u^{(\gamma)}_{p_2}(\tau')u^{(\gamma)}_{p_3}(\tau')}{u^{(\gamma)}_{p_1}(\tau)u^{(\gamma)}_{p_2}(\tau)u^{(\gamma)}_{p_3}(\tau)} \nonumber \\
	&=-\frac{i }{H^2 \left(i+\tau  p_1\right) \left(i+\tau  p_2\right)
		\left(i+\tau  p_3\right)} \nonumber \\
	&\times \left[\frac{i}{\tau }+\frac{p_1^2 \left(p_2+p_3\right)+p_2 p_3 \left(p_2+p_3\right)+p_1
		\left(p_2^2+4 p_2 p_3+p_3^2\right)}{\left(p_1+p_2+p_3\right){}^2}-\frac{i \tau  p_1 p_2 p_3}{p_1+p_2+p_3}\right]\nonumber \\
	&\approx-\frac{i a(\tau)}{H}-\frac{p_1^3+p_2^3+p_3^3+2 \left(p_2+p_3\right) p_1^2+2 \left(p_2^2+p_3^2\right) p_1+2 p_2 p_3^2+2 p_2^2 p_3+2p_1p_2p_3}{H^2
		\left(p_1+p_2+p_3\right){}^2} \nonumber \\
	& +\mathcal{O}(\tau) \ , \label{eq:I_gamma_gamma_gamma}
\end{align}
and plugging the leading imaginary part into (\ref{eq:NG_3gamma}) gives the growing phase for the wave functional
\begin{align}
	\Psi_{NG}^{(\gamma\gamma\gamma)}&\approx \exp\left[i\frac{a M_p^2}{8H}\int d^3x \left(2\partial_m\gamma {}_i{}_l \partial_l\gamma {}_j{}_m \gamma 
	{}_i{}_j+  \partial_i\gamma {}_l{}_m \partial_j\gamma {}_l{}_m \gamma{}_i{}_j\right)+\mathcal{O}(a^0)\right] \ , \label{eq:NG_3gamma_late_time}
\end{align}
which agrees with the second line of (\ref{eq:W3_cubic_HJ}). It is noteworthy that the integral (\ref{eq:I_gamma_gamma_gamma}) is also discussed in \cite{Nelson:2016kjm} for the scalar cubic interaction proportional to $a\zeta\left(\partial_i\zeta\right)^2$,\footnote{Except the overall factor involving polarization tensors and comoving momenta, the bulk interactions with the form $a\zeta\left(\partial\zeta\right)^2$ and $a\gamma\left(\partial\gamma\right)^2$ leads the same time integral (\ref{eq:I_gamma_gamma_gamma}).} in which the corresponding oscillating phase is the main contribution to the decoherence of $\zeta$, and we will calculate the similar decoherence for $\mathcal{L}_{\gamma\gamma\gamma}$ in Sec. \ref{sec:3gamma_decohernce}.

On the other hand, the last boundary term with four spatial derivatives in the first line of $\mathcal{L}_{{\rm bd},\zeta\zeta\zeta}$ (\ref{eq:3zeta_bd}) is included in the next order solution of the HJ equation, which has been derived in \cite{Parry:1993mw}\footnote{In \cite{Parry:1993mw}, the solution is calculated with the explicit potential $V(\phi)=V_0 \exp\left(-\sqrt{\frac{2}{p}}\frac{\phi}{M_p}\right)$, and the terms depending on the parameter $p$ can be easily rewritten in terms of the slow-roll parameter with $p\approx\frac{1}{\epsilon}$.}
\begin{align}
	\Delta W (h_{ij})&=\frac{M_p^2}{2}\int_{\Sigma}d^3x \frac{\sqrt{h}}{H^3}\left({}^{(3)}R_{ij}{}^{(3)}R^{ij} -\frac{3}{8}{}^{(3)}R^2\right) +\mathcal{O}(\epsilon,\eta) \nonumber \\
	&\supset -M_p^2\int_\Sigma d^3x\frac{1}{4aH^3}\left(\partial_i\zeta \right)^2 \partial^2\zeta \ .
\end{align}
where we only keep the slow-roll unsuppressed terms with the $\zeta$-gauge ($\partial_i\phi=0$) in the first line. Since the terms with four spatial derivatives are suppressed by the scale factor as $a^3\frac{\partial^4}{a^4}=\frac{\partial^4}{a}$, we expect that their contribution to decoherence is negligible at late time.

We emphasize that the cubic WKB phase (\ref{eq:W3_cubic_HJ}) is independent to how the integration by parts is chosen to split bulk and boundary interaction terms in (\ref{eq:splitting_bulk_boundary}), as supported by the appearance of $\gamma\gamma\gamma$ terms, and it relies on the hypersurface $\Sigma$ for evaluating the WDW wave functional, as discussed in \cite{Sou:2022nsd} by comparing the boundary terms on different hypersurfaces. Therefore, the wave functional of cosmological perturbations is expected to have slow-roll unsuppressed non-Gaussian phase regardless of the way of doing integration by parts in the action (\ref{eq:action_single_field}), and all the three methods presented here give consistent results as long as the slow-roll unsuppressed terms are identified.

\section{The decoherence of primordial gravitons} \label{eq:decoherence}
In this section, we calculate the decoherence with the wave functional of cosmological perturbations $\Psi(\zeta,\gamma_{ij})$ and the formalism used in \cite{Nelson:2016kjm,Liu:2016aaf,Sou:2022nsd}. We consider the primordial gravitons to be observed form a system $\{\xi_{\bf q}\}=\{\gamma_{\bf q}^s\}$, and other unobserved degrees of freedom interacting with the system form the environment $\{\mathcal{E}_{\bf k}\}$, which includes unobserved modes of scalar $\zeta_{\bf k}$ and tensor perturbations $\gamma_{\bf k}^s$. With the cubic scalar-tensor and three-tensor interactions, the states of gravitons' modes cannot evolve independently, and they entangle with the environment, represented by the non-Gaussian part of the wave functional
\begin{align}
	\Psi(\zeta,\gamma_{ij})&=\Psi_G(\zeta,\gamma)\Psi_{NG}(\zeta,\gamma) \nonumber \\
	&\approx\Psi^{(\xi)}_G(\xi)\Psi^{(\mathcal{E})}_G(\mathcal{E})\exp\left(\sum_{\sigma_1,\sigma_2,s}\int_{{\bf k},{\bf k}',{\bf q}}\mathcal{F}^{\sigma_1,\sigma_2,s}_{{\bf k},{\bf k}',{\bf q}}\mathcal{E}^{\sigma_1}_{\bf k}\mathcal{E}^{\sigma_2}_{{\bf k}'}\xi^{s}_{\bf q}\right) \ ,
\end{align}
where $\sigma_i$ denotes some discrete degrees of freedom of the environment modes, two polarizations for tensor and one mode for scalar perturbations, and we focus on the non-Gaussian part involving two environment and one system modes since this dominates the decoherence \cite{Burgess:2022nwu}.

As the state of the environment cannot be accessed, the system is described by the reduced density matrix obtained by tracing out unobserved degrees of freedom
\begin{align}
	\rho_R(\xi,\tilde{\xi})&=\Psi^{(\xi)}_G(\xi)\left[\Psi^{(\xi)}_G(\tilde\xi)\right]^* \left\langle \exp\left[\sum_{\sigma_1,\sigma_2,s}\int_{{\bf k},{\bf k}',{\bf q}}\mathcal{E}^{\sigma_1}_{\bf k}\mathcal{E}^{\sigma_2}_{{\bf k}'}\left(\mathcal{F}^{\sigma_1,\sigma_2,s}_{{\bf k},{\bf k}',{\bf q}}\xi^{s}_{\bf q}+{\mathcal{F}^*}^{\sigma_1,\sigma_2,s}_{{\bf k},{\bf k}',{\bf q}}\tilde\xi^{s}_{\bf q}\right)\right]  \right\rangle_{\mathcal{E}} \ ,
\end{align}
where $\langle \cdots \rangle_{\mathcal{E}}=\int D\mathcal{E}\left|\Psi^{(\mathcal{E})}_G(\mathcal{E})\right|^2\left(\cdots\right)$ (similar notation $\langle \cdots \rangle_{\xi}$ for the system). In the cases when the non-Gaussian part is a rapidly oscillating phase (with $\mathcal{F}^{\sigma_1,\sigma_2,s}_{{\bf k},{\bf k}',{\bf q}}$ has a dominated imaginary part), we expect that the expectation value over environment modes is highly suppressed if $\xi\neq\tilde{\xi}$, characterizing the loss of interference (decoherence). Such a suppression of off-diagonal terms of $\rho_R$ is calculated as the decoherence factor
\begin{align}
	D(\xi,\tilde{\xi})&=\left|\frac{\rho_R(\xi,\tilde{\xi})}{\sqrt{\rho_R(\xi,\xi)\rho_R(\tilde{\xi},\tilde{\xi})}}\right| \ ,
\end{align}
and for a particular system mode with comoving momentum ${\bf q}$, the leading contribution is a one-loop integral\footnote{Here we mean the integral can be interpreted as a one-loop diagram, and one should not be confused with the one-loop quantum correction discussed in Sec. \ref{sec:1_loop_correction}.}
\begin{align}
	&D(\xi_{\bf q},\tilde{\xi}_{\bf q}) \nonumber \\
	&\approx \exp\Bigg[-\frac{1}{2V}\sum_{\sigma_1,\cdots,\sigma_4,s,s'}\int_{{\bf k}+{\bf k}'=-{\bf q}} {\rm Im}\mathcal{F}^{\sigma_1,\sigma_2,s}_{{\bf k},{\bf k}',{\bf q}}{\rm Im}\mathcal{F}^{\sigma_3,\sigma_4,s'}_{-{\bf k},-{\bf k}',-{\bf q}} \left(\langle \mathcal{E}_{\bf k}^{\sigma_1}\mathcal{E}_{-{\bf k}}^{\sigma_3}\rangle_{\mathcal{E}}'\langle \mathcal{E}_{{\bf k}'}^{\sigma_2}\mathcal{E}_{-{\bf k}'}^{\sigma_4}\rangle_{\mathcal{E}}'+1 \ {\rm perm}\right) \nonumber \\
	&\times
	 (\xi^s_{\bf q}-\tilde{\xi}^s_{\bf q})(\xi^{s
		'}_{-{\bf q}}-\tilde{\xi}^{s'}_{-{\bf q}})\Bigg]  \ ,
\end{align}
where the prime $\langle \cdots \rangle_{\mathcal{E}}'$ means ignoring factors like $(2\pi)^3\delta^3({\bf k}_1+{\bf k}_2)$, and  the volume $V=(2\pi)^3\delta^3({\bf 0})$ is for discretizing the integral $\int\frac{d^3q}{(2\pi)^3}\to \frac{\sum_{\bf q}}{V}$. We only keep the imaginary part of $\mathcal{F}^{\sigma_1,\sigma_2,s}_{{\bf k},{\bf k}',{\bf q}}$ which dominates in all the cases studied in this paper, and the expectation value of the minus exponent defines the dimensionless  "decoherence exponent" (initially called the "decoherence rate" in \cite{Nelson:2016kjm})\footnote{Some papers \cite{Gong:2019yyz,Burgess:2022nwu} define the physical decoherence rate in the usual sense, so it is better to use another name for avoiding ambiguity while comparing some results. Note that we can also define the physical decoherence rate $\frac{d}{dt}\Gamma(q,\tau)=\mathcal{O}(1)H\Gamma(q,\tau)$ since $\Gamma(q,\tau)\propto a^n$, and the decoherence moment with $\Gamma(q,\tau)\approx 1$ means the physical decoherence rate is comparable to the Hubble rate.}
\begin{align}
	\Gamma(q,\tau) &=P_q^{(\gamma)}\sum_{\sigma_1,\cdots,\sigma_4,s}\int_{{\bf k}+{\bf k}'=-{\bf q}} {\rm Im}\mathcal{F}^{\sigma_1,\sigma_2,s}_{{\bf k},{\bf k}',{\bf q}}{\rm Im}\mathcal{F}^{\sigma_3,\sigma_4,s}_{-{\bf k},-{\bf k}',-{\bf q}} \left(\langle \mathcal{E}_{\bf k}^{\sigma_1}\mathcal{E}_{-{\bf k}}^{\sigma_3}\rangle_{\mathcal{E}}'\langle \mathcal{E}_{{\bf k}'}^{\sigma_2}\mathcal{E}_{-{\bf k}'}^{\sigma_4}\rangle_{\mathcal{E}}'+1 \ {\rm perm}\right) \ ,
\end{align}
where we use $\left\langle (\xi^s_{\bf q}-\tilde{\xi}^s_{\bf q})(\xi^{s
	'}_{-{\bf q}}-\tilde{\xi}^{s'}_{-{\bf q}})\right\rangle_\xi=2P^{(\gamma)}_qV\delta_{s,s'}$, and the decoherence of the system mode $\xi_{\bf q}$ happens when $\Gamma(q,\tau)\approx 1$. We will study the decoherence of primordial gravitons by the two boundary terms in $\mathcal{L}_{{\rm bd},\zeta-\gamma}$ (\ref{eq:scalar_tensor_boundary}) and the bulk term $\mathcal{L}_{\gamma\gamma\gamma}$ (\ref{eq:3gamma_bulk}), which include three leading terms decohering $\gamma_{ij}$ by the orders of magnitudes in Table \ref{tab:int_terms},\footnote{The decoherence by the bulk term $\epsilon a \gamma_{ij}\partial_i\zeta \partial_j\zeta$ has been calculated in \cite{Burgess:2022nwu}, so we do not repeat the calculation here.} and the corresponding $\Gamma(q,\tau)$ are computed as the diagrams in Fig. \ref{fig:diagrams_gamma_deco}.
\begin{figure}
	\centering
\tikzset{every picture/.style={line width=0.75pt}} 
\begin{tikzpicture}[x=0.75pt,y=0.75pt,yscale=-1.3,xscale=1.3]
		 
		\draw [line width=1.5, snake it]    (26,80.23) -- (70.74,80.23) ;
		\draw  [color={rgb, 255:red, 255; green, 0; blue, 0 }  ,draw opacity=1 ][line width=1.5]  (70.74,80.23) .. controls (70.74,67.88) and (80.76,57.86) .. (93.12,57.86) .. controls (105.47,57.86) and (115.49,67.88) .. (115.49,80.23) .. controls (115.49,92.59) and (105.47,102.6) .. (93.12,102.6) .. controls (80.76,102.6) and (70.74,92.59) .. (70.74,80.23) -- cycle ;
		\draw [line width=1.5, snake it]    (115.49,80.23) -- (160.23,80.23) ;
		
		\draw [line width=1.5, snake it]    (183,80.23) -- (227.74,80.23) ;
		\draw [line width=1.5, snake it]    (272.49,80.23) -- (317.23,80.23) ;
		\draw  [color={rgb, 255:red, 255; green, 0; blue, 0 }  ,draw opacity=1][line width=1.5]  (227.74,80.23) .. controls (227.74,67.88) and (237.76,57.86) .. (250.12,57.86) .. controls (262.47,57.86) and (272.49,67.88) .. (272.49,80.23)  ;
		\draw  [color={rgb, 255:red, 255; green, 0; blue, 0 }  ,draw opacity=1] [line width=1.5, snake it,shorten >= 3pt] (272.49,80.23) .. controls (272.49,92.59) and (262.47,102.6) .. (250.12,102.6) .. controls (237.76,102.6) and (227.74,92.59) .. (225.5,77.9) ;

		\draw [line width=1.5, snake it]    (26+314,80.23) -- (70.74+314,80.23) ;
		\draw  [color={rgb, 255:red, 255; green, 0; blue, 0 }  ,draw opacity=1 ][line width=1.5, snake it]  (70.74+314,80.23) .. controls (70.74+314,67.88) and (80.76+314,57.86) .. (93.12+314,57.86) .. controls (105.47+314,57.86) and (115.49+314,67.88) .. (115.49+314,80.23) .. controls (115.49+314,92.59) and (105.47+314,102.6) .. (93.12+314,102.6) .. controls (80.76+314,102.6) and (70.74+314,92.59) .. (70.74+314,80.23) -- cycle ;
		\draw [line width=1.5, snake it]    (115.49+314,80.23) -- (160.23+314,80.23) ;

		\draw (41.77,85.1) node [anchor=north west][inner sep=0.75pt]    {$q$};
		\draw (131.26,85.1) node [anchor=north west][inner sep=0.75pt]    {$q$};
		\draw (83.62,32.7) node [anchor=north west][inner sep=0.75pt]  [color={rgb, 255:red, 255; green, 0; blue, 0 }  ,opacity=1 ]  {$P^{(\zeta)}_{k}$};
		\draw (84.62,108.7) node [anchor=north west][inner sep=0.75pt]  [color={rgb, 255:red, 255; green, 0; blue, 0 }  ,opacity=1 ]  {$P^{(\zeta)}_{k'}$};
		
		\draw (198.77,85.1) node [anchor=north west][inner sep=0.75pt]    {$q$};
		\draw (288.26,85.1) node [anchor=north west][inner sep=0.75pt]    {$q$};
		\draw (240.62,32.7) node [anchor=north west][inner sep=0.75pt]  [color={rgb, 255:red, 255; green, 0; blue, 0 }  ,opacity=1 ]  {$P^{(\zeta)}_{k}$};
		\draw (241.62,108.7) node [anchor=north west][inner sep=0.75pt]  [color={rgb, 255:red, 255; green, 0; blue, 0 }  ,opacity=1 ]  {$P^{(\gamma)}_{k'}$};
		
		\draw (41.77+314,85.1) node [anchor=north west][inner sep=0.75pt]    {$q$};
		\draw (131.26+314,85.1) node [anchor=north west][inner sep=0.75pt]    {$q$};
		\draw (83.62+314,32.7) node [anchor=north west][inner sep=0.75pt]  [color={rgb, 255:red, 255; green, 0; blue, 0 }  ,opacity=1 ]  {$P^{(\gamma)}_{k}$};
		\draw (84.62+314,108.7) node [anchor=north west][inner sep=0.75pt]  [color={rgb, 255:red, 255; green, 0; blue, 0 }  ,opacity=1 ]  {$P^{(\gamma)}_{k'}$};
	\end{tikzpicture}
	\caption{The diagrams for computing $\Gamma(q,\tau)$. The system and environment are labeled with black and red colors respectively, and the scalar and tensor perturbations are denoted by straight and wavy lines respectively. \label{fig:diagrams_gamma_deco}}
\end{figure}

\subsection{$\zeta\zeta\gamma$ boundary interaction}
With the $\zeta\zeta\gamma$ boundary term in (\ref{eq:scalar_tensor_boundary}), the wavefunctional has a non-Gaussian phase
\begin{align}
	\Psi^{(\zeta\zeta\gamma)}_{NG}(\zeta,\gamma)=\exp\left(\sum_{s}i\frac{M_p^2}{H}a\int_{{\bf k},{\bf k}',{\bf q}}k_i k'_j e^s_{ij}({\bf q}) \zeta_{\bf k}\zeta_{{\bf k}'}\gamma^s_{\bf q}\right) \ ,
\end{align}
leading to the decoherence factor for the tensor mode with comoving momentum ${\bf q}$
\begin{align}
	&D^{\rm bd}_{\zeta\zeta\gamma}(\gamma_{\bf q},\tilde{\gamma}_{\bf q}) \nonumber \\
	&\approx \exp\left[-\sum_{s,s'}\frac{M_p^4a^2}{H^2}\frac{1}{V}\int_{{\bf k}+{\bf k}'=-{\bf q}}P_k^{(\zeta)}P_{k'}^{(\zeta)}k_i{k'_j} e^{s}_{ij}({\bf q})k_lk'_m e^{s'}_{lm}(-{\bf q})(\gamma^s_{\bf q}-\tilde{\gamma}^s_{\bf q})(\gamma^{s
		'}_{-{\bf q}}-\tilde{\gamma}^{s'}_{-{\bf q}})\right] , \label{eq:Deco_factor_zeta_zeta_gamma}
\end{align}
and the decoherence exponent
\begin{align}
	\Gamma^{\rm bd}_{\zeta\zeta\gamma}(q,\tau)
	&= \frac{a^2M_p^2}{q^3} \sum_{s}\int_{{\bf k}+{\bf k}'=-{\bf q}}P_k^{(\zeta)}P_{k'}^{(\zeta)}k_ik'_j e^s_{ij}({\bf q})k_lk'_me^s_{lm}(-{\bf q})\label{eq:Gamma_zeta_zeta_gamma} \ .
\end{align}
Here we want to study the decoherence of the tensor mode by both the sub- and super-horizon scalar environments, the integral over ${{\bf k},{\bf k}'}$ should be all the scalar modes except the observable super-horizon modes denoted as $q_{\zeta,{\rm min}} \le k,k'\le q_{\zeta,{\rm max}}$. We will show later with the explicit result that such an exclusion of the observable region only modifies the sub-dominated part of the decoherence, so we first calculate (\ref{eq:Gamma_zeta_zeta_gamma}) with all $0<k,k'<+\infty$.

To calculate (\ref{eq:Gamma_zeta_zeta_gamma}), we can choose $\bf q$ on the z-axis, and the polarization tensors along the $\bf q$-axis are
\begin{align}
	e^+_{ij}({\bf q})=\begin{pmatrix}
		1 & i & 0 \\
		i & -1 & 0 \\
		0 & 0 & 0 
	\end{pmatrix} \ , \ e^-_{ij}({\bf q})=\begin{pmatrix}
		1 & -i & 0 \\
		-i & -1 & 0 \\
		0 & 0 & 0  
	\end{pmatrix} \ ,
\end{align}
with $e^{\pm}_{ij}(-{\bf q})=\left(e^{\pm}_{ij}({\bf q})\right)^*$,
so the multiplication of comoving momenta and polarization tensors is
\begin{align}
	\sum_{s}k_ik'_j e^s_{ij}({\bf q})k_lk'_me^s_{lm}(-{\bf q}) 
	&=2(k_x^2+k_y^2)({k'}_x^2+{k'}_y^2) \nonumber \\
	&=2k^4(1-\cos^2\theta)^2 \nonumber \\
	&=2k^4 \sin^4\theta \ , \label{eq:product_polarization_1}
\end{align} 
where here $k_z=k\cos \theta$, and we applied the fact that $|k_x|=|k'_x|$ and $|k_y|=|k'_y|$ since ${\bf k}+{\bf k}'=-q \hat{\bf z}$. Therefore, the conserved-momentum integral in (\ref{eq:Deco_factor_zeta_zeta_gamma}) is calculated with the spherical coordinates
\begin{align}
	\Gamma^{\rm bd}_{\zeta\zeta\gamma}(q,\tau)
	&=\frac{2a^2M_p^2}{q^3}\int_{k<a\Lambda}\frac{d^3k}{(2\pi)^3}P_k^{(\zeta)}P_{\sqrt{k^2+q^2-2kq\cos\theta}}^{(\zeta)}k^4\sin^4\theta \nonumber \\
	&=\frac{2a^2M_p^2}{q^3}\left(\frac{H^2}{4M_p^2\epsilon}\right)^2\Bigg[-\frac{4 \Lambda ^5}{75 \pi ^2 H^5 \tau }+\frac{8 \Lambda ^3 \left(q^2 \tau ^2-7\right)}{315 \pi ^2 H^3 \tau }-\frac{4 \Lambda  \left(q^4
		\tau ^4-15 q^2 \tau ^2+21\right)}{315 \pi ^2 H \tau } \nonumber \\
	&-\frac{q \left(q^4 \tau ^4-25 q^2 \tau ^2+75\right)}{225 \pi ^2}\Bigg] \ , \label{eq:power-law_divergence}
\end{align}
with a UV cutoff $a\Lambda$. The result has a few power-law UV divergences but without logarithmic type, so using the dimensional regularization (dim. reg.) only keeps the last term in (\ref{eq:power-law_divergence}), which is negative. We follow the method used in \cite{Sou:2022nsd} resolving the UV divergence for the $\zeta$ boundary term (\ref{eq:Lbd_zeta_previous}), and this involves a field redefinition of $\zeta$
\begin{align}
	\zeta_{\bf p}\to \bar{\zeta}_{\bf p}=\left(1+\frac{p^2}{a^2H^2}\right)\zeta_{\bf p} \ , \label{eq:zeta_field_redefinition}
\end{align}
which dominantly redefines the sub-horizon modes, whereas the change of super-horizon modes is suppressed for preserving the correct nearly scale-invariant power spectrum. Note that the similar idea of resolving the UV divergence by field redefinition is also demonstrated in \cite{Barvinsky:1998cq} for the decoherence of background scale factor $a(t)$ by scalar field (with the setup in a closed spacetime), where the failure of using local counterterms and the dim. reg. is discussed.

With (\ref{eq:zeta_field_redefinition}), the scalar power spectrum is changed to
\begin{align}
	P^{(\zeta)}_p(\tau)=\frac{H^2}{4\epsilon M_p^2}\frac{1}{p^3 \left(1+p^2\tau^2\right)} \ ,
\end{align}
and the integrand of (\ref{eq:power-law_divergence}) scales as $k^{-6}$ when $k\to+\infty$, causing the integral to converge:
\begin{align}
	\Gamma^{\rm bd}_{\zeta\zeta\gamma}(q,\tau)&=-\frac{2a^2M_p^2}{q^3\tau}\left(\frac{H^2}{4M_p^2\epsilon}\right)^2 J_{\zeta\zeta\gamma}\left(\frac{q}{aH}\right)\nonumber \\
	&\approx \frac{\pi \Delta_\zeta^2}{15\epsilon}\left(\frac{aH}{q}\right)^3 +\mathcal{O}(a^2) \ , \label{eq:Gamma_zeta_zeta_gamma_renormalized}
\end{align}
where the full analytical expression of function $J_{\zeta\zeta\gamma}\left(Q\right)$ is shown in Appendix (\ref{eq:explicit_J_zeta_zeta_gamma}).

Finally, we justify the previous claim that the exclusion of observable super-horizon region  $q_{\zeta,{\rm min}} \le k,k'\le q_{\zeta,{\rm max}}$ is sub-dominated as follows. The power spectrum of these modes converges as $P^{(\zeta)}_{k}\propto \frac{1}{k^3}+\mathcal{O}(\left(k\tau\right)^2)$, so their contribution to the integrand in (\ref{eq:power-law_divergence}) converges to a constant at late time, corresponding to the $\mathcal{O}(a^2)$ contribution in the decoherence exponent. Similar fact is also reported in \cite{Burgess:2022nwu} with the Lindblad equation approach that the dependence on the partition of system and environment is in the sub-dominated order at late time.

\subsection{$\zeta\gamma\gamma$ boundary interaction}
Now we consider the second term of (\ref{eq:scalar_tensor_boundary}), and we split the tensor modes into observable (system $\xi_{ij}$) and unobservable (environment $\mathcal{E}_{ij}$) by their comoving momenta:
\begin{align}
	\gamma_{ij}({\bf x},t)=\sum_{s}\left[\int\frac{d^3q}{(2\pi)^3} \xi^s_{\bf q}(t)e^s_{ij}({\bf q})e^{i{\bf q}\cdot {\bf x}}+\int \frac{d^3k}{(2\pi)^3}\mathcal{E}^s_{\bf k}(t)e^s_{ij}({\bf k})e^{i{\bf k}\cdot {\bf x}}\right] \ , \label{eq:gamma_split_system_environment}
\end{align}
and the wave functional includes a phase 
\begin{align}
	\Psi^{(\zeta\gamma\gamma)}_{NG}(\zeta,\gamma)&=\exp\left(\sum_{s,s'}i\frac{M_p^2}{8H}a\int_{{\bf p},{\bf p}',{\bf p}^{''}} \left({{\bf p}'}\cdot{{\bf p}''}\right) e^s_{ij}({\bf p}'') e^{s'}_{ij}({{\bf p}'}) \zeta_{\bf p}\gamma^{s'}_{{\bf p}'}\gamma^s_{{\bf p}''}\right) \nonumber \\
	&\supset\exp\left(\sum_{s,s'}i\frac{M_p^2}{4H}a\int_{{\bf k},{\bf k}',{\bf q} }\left({\bf k}'\cdot {\bf q}\right) e^s_{ij}({\bf q}) e^{s'}_{ij}({{\bf k}'}) \zeta_{\bf k}\mathcal{E}^{s'}_{{\bf k}'}\xi^s_{\bf q}\right) \ .
\end{align}
The decoherence factor and exponent are 
\begin{align}
	&D^{\rm bd}_{\zeta\gamma\gamma}(\xi_{\bf q},\tilde{\xi}_{\bf q})\nonumber \\
	&\approx \exp\Bigg[-\sum_{s_1,s_2,s_3}\frac{1}{2V}\left(\frac{aM_p^2}{4H}\right)^2\int_{{\bf k}+{\bf k}'=-{\bf q}} \left({\bf k}'\cdot {\bf q}\right)^2 e^{s_1}_{ij}({\bf q})e^{s_2}_{lm}(-{\bf q}) e^{s_3}_{ij}({{\bf k}'})  e^{s_3}_{lm}({-{\bf k}'})P_k^{(\zeta)}P_{k'}^{(\gamma)} \nonumber \\
	&\times(\xi^{s_1}_{\bf q}-\tilde{\xi}^{s_1}_{\bf q})(\xi^{s
		_2}_{-{\bf q}}-\tilde{\xi}^{s_2}_{-{\bf q}})\Bigg] \ ,
\end{align}
and 
\begin{align}
	\Gamma^{\rm bd}_{\zeta\gamma\gamma}(q,\tau) 
	&=\frac{M_p^2}{32}\frac{a^2}{q^3}\sum_{s,s'}\int_{{\bf k}+{\bf k}'=-{\bf q}} \left({\bf k}'\cdot {\bf q}\right)^2 e^s_{ij}({\bf q})  e^s_{lm}(-{\bf q})e^{s'}_{ij}({{\bf k}'}) e^{s'}_{lm}({-{\bf k}'})P_k^{(\zeta)}P_{k'}^{(\gamma)}  \ , \label{eq:Gamma_zeta_gamma_gamma_express}
\end{align}
respectively. To compute the product of the four polarization tensors, we express the polarization tensor with ${\bf k}'=k'\left(\sin\theta\cos\phi,\sin\theta\sin\phi,-\cos\theta\right)$ in the coordinate form, and this has been done, e.g. in the appendix of \cite{Gong:2019yyz}\footnote{We set the coordinates such that ${\bf k}'=-{\bf q}$ when $\theta=0$, and our convention of the polarization tensors $e^{\pm}_{ij}$ is related to the one used in \cite{Gong:2019yyz} $\tilde{e}^{+,\times}_{ij}$ by the linear combinations: $e^\pm_{ij}=\tilde{e}^{+}_{ij}\pm i\tilde{e}^{\times}_{ij}$.}
\begin{align}
	e^{\pm}_{ij}({\bf k}')&=\left(\hat{\bf e}_1\right)_i\left(\hat{\bf e}_1\right)_j-\left(\hat{\bf e}_2\right)_i\left(\hat{\bf e}_2\right)_j\pm i\left(\hat{\bf e}_1\right)_i\left(\hat{\bf e}_2\right)_j+\pm i\left(\hat{\bf e}_2\right)_i\left(\hat{\bf e}_1\right)_j \ ,
\end{align}
where
\begin{align}
	\hat{\bf e}_1&=\left(-\cos\theta\cos\phi,-\cos\theta\sin\phi,-\sin\theta\right) \nonumber \\
	\hat{\bf e}_2&=(-\sin\phi,\cos\phi,0) \ .
\end{align}
With this, the product of the four polarization tensors in (\ref{eq:Gamma_zeta_gamma_gamma_express}) becomes
\begin{align}
	\sum_{s,s'}e^s_{ij}({\bf q})  e^s_{lm}(-{\bf q})e^{s'}_{ij}({{\bf k}'}) e^{s'}_{lm}({-{\bf k}'})&=32\left(\cos^8\frac{\theta}{2}+\sin^8\frac{\theta}{2}\right) \ , \label{eq:product_polarization_2}
\end{align}
and plugging in to (\ref{eq:Gamma_zeta_gamma_gamma_express}) becomes
\begin{align}
	&\Gamma^{\rm bd}_{\zeta\gamma\gamma}(q,\tau) \nonumber \\
	&=M_p^2\frac{a^2}{q^3}\int_{k'<a\Lambda,\ |k'-q|>k_{\rm min}}\frac{d^3k'}{(2\pi)^3} k'^2 q^2\cos^2\theta\left(\cos^8\frac{\theta}{2}+\sin^8\frac{\theta}{2}\right) P_{\sqrt{k'^2+q^2-2k'q\cos\theta}}^{(\zeta)}P_{k'}^{(\gamma)} \nonumber \\
	&=-\frac{H^4}{64M_p^2\epsilon}\frac{a^2}{q^2}\Bigg[\frac{88 \Lambda ^3 q \tau }{315 \pi ^2 H^3}+\frac{16 \Lambda  q \tau  \left(3 q^2 \tau ^2+11\right)}{105 \pi ^2 H}-\frac{4 \left(q^2 \tau ^2+1\right) \log \left(\frac{q}{k_{\min }}\right)}{\pi ^2} \nonumber \\
	&+\frac{31 q^4 \tau ^4+402 q^2 \tau ^2+474}{90 \pi ^2}\Bigg] \ , \label{eq:Gamma_zeta_gamma_gamma_integral}
\end{align}
which has UV and IR divergences. The UV divergence can be resolved by the field redefinition (\ref{eq:zeta_field_redefinition}) and a similar one for $\gamma_{ij}$, making the integral (\ref{eq:Gamma_zeta_gamma_gamma_integral}) converges at $k'\to+\infty$
\begin{align}
	\Gamma^{\rm bd}_{\zeta\gamma\gamma}(q,\tau)&=\frac{H^4}{64M_p^2\epsilon}\frac{a^2}{q^2}\left[\frac{4}{\pi^2\left(1+q^2\tau^2\right)}\log\left(\frac{q}{k_{\rm min}}\right)+J_{\zeta\gamma\gamma}\left(\frac{q}{aH}\right)\right] \nonumber \\
	&\approx\frac{\Delta^2_\zeta}{120}\left[60\log\left(\frac{q}{k_{\rm min}}\right)-79\right]\left(\frac{aH}{q}\right)^2 +\mathcal{O}\left(a\right) \ , \label{eq:Gamma_zeta_gamma_gamma_renormalized}
\end{align}
where the analytical expression of $J_{\zeta\gamma\gamma}\left(Q\right)$ is shown in Appendix (\ref{eq:explicit_J_zeta_gamma_gamma}). The IR divergence can be regularized by putting the finite duration of inflation as a cutoff with $\log\left(\frac{q}{k_{\rm min}}\right)=N_q-N_{\rm IR}=\Delta N$, counting the e-folds from the onset of inflation to the horizon crossing, and this cutoff has been applied in \cite{Martin:2018zbe}. Similar to the case of $\zeta\zeta\gamma$, the exclusion of some observable super-horizon modes also contributes to $\mathcal{O}(a^2)$ term to (\ref{eq:Gamma_zeta_gamma_gamma_renormalized}), which is negligible compared to the logarithmic factor. On the hand, the lack of $\mathcal{O}(a^2)$ IR divergence in the case of $\zeta\zeta\gamma$ is manifested by comparing the products of polarization tensors in the limit ${\bf k}\to -{\bf q}$ (or $\theta\to 0$) in (\ref{eq:product_polarization_1}) and (\ref{eq:product_polarization_2}), and the former has vanishing contribution to the decoherence exponent, whereas the latter contributes.

\subsection{$\gamma\gamma\gamma$ bulk interaction} \label{sec:3gamma_decohernce}
We note that the calculations of three-tensor decoherence in the literature are done with $h_{ij}^{TT}$ \cite{Gong:2019yyz},\footnote{In \cite{Burgess:2022nwu}, the authors mentioned that the full calculation of the three-tensor decoherence (with $\gamma_{ij}$) with the Lindblad equation is in preparation, and here we just give an estimation of how large it is.} and we want to see if the difference between this and $\gamma_{ij}$ discussed in Sec. \ref{sec:setup} leads to deviations of decoherence exponent. As shown in (\ref{eq:I_gamma_gamma_gamma}), the non-Gaussian phase contributed by $\mathcal{L}_{\gamma\gamma\gamma}$ has the same time-dependent structure as the three-scalar interaction $a\zeta\left(\partial_i\zeta\right)^2$, so we can generalize the calculation in \cite{Nelson:2016kjm} to the three-tensor case. We apply the same squeezed limit $k\approx k'\gg q$ and ${\bf k}\approx -{\bf k}'$ to estimate the dominated part of the decoherence, and it turns out can greatly simplify the product of polarization tensors:
\begin{align}
	&\mathcal{L}_{\gamma\gamma\gamma} \nonumber \\
	&\to -\frac{M_p^2}{8}a\sum_{s_1,s_2,s_3}\left[2k_mk'_le^{s_1}_{il}({\bf k})e^{s_2}_{jm}({\bf k}')+k_ik'_je^{s_1}_{lm}({\bf k})e^{s_2}_{lm}({\bf k}')\right]e^{s_3}_{ij}({\bf q})\mathcal{E}^{s_1}_{{\bf k}}\mathcal{E}^{s_2}_{{\bf k}'}\xi^{s_3}_{{\bf q}}+\mathcal{O}(kq,k'q) \nonumber \\
	&\approx \frac{M_p^2}{2}a\sum_{s_1,s_3} k_i k_je^{s_3}_{ij}({\bf q})\mathcal{E}^{s_1}_{{\bf k}}\mathcal{E}^{s_1}_{{\bf k}'}\xi^{s_3}_{{\bf q}}+\mathcal{O}(kq,k'q) \ , \label{eq:squeezed_limit_3gamma}
\end{align}
where in the last line we use the facts $k_m e^{s_2}_{jm}({\bf k}')\approx-k'_m e^{s_2}_{jm}({\bf k}')=0$ and $e^{s_1}_{lm}({\bf k})e^{s_2}_{lm}({\bf k}')\approx e^{s_1}_{lm}({\bf k})e^{s_2}_{lm}(-{\bf k})=4\delta_{s_1,s_2}$, and the factor of polarization tensors agrees with the graviton's consistency relation \cite{Maldacena:2002vr}. With (\ref{eq:NG_3gamma}), the dominated phase part of (\ref{eq:I_gamma_gamma_gamma}) and (\ref{eq:squeezed_limit_3gamma}), the decoherence factor is approximated as
\begin{align}
	D^{\rm sq}_{\gamma\gamma\gamma}(\xi_{\bf q},\tilde{\xi}_{\bf q}) 
	&\approx \exp\Bigg[-\frac{M_p^4}{2V}\sum_{s,s'}\int_{{\bf k}+{\bf k}'=-{\bf q}}^{k\approx k'\gg q}k_i k'_je^{s}_{ij}({\bf q})k_l k'_me^{s'}_{lm}(-{\bf q})\left({\rm Im}I^{(\gamma\gamma\gamma)}_{{\bf k},{\bf k}',{\bf q}}\right)^2 P_{k}^{(\gamma)}P_{k'}^{(\gamma)} \nonumber \\
	&\times(\xi^s_{\bf q}-\tilde{\xi}^s_{\bf q})(\xi^{s
		'}_{-{\bf q}}-\tilde{\xi}^{s'}_{-{\bf q}})\Bigg] \ ,
\end{align}
and the decoherence exponent is
\begin{align}
	\Gamma^{\rm sq}_{\gamma\gamma\gamma}(q,\tau)&\approx \frac{H^2M_p^2}{q^3}\int\frac{d^3k}{(2\pi)^3} k^4\left({\rm Im}I^{(\gamma\gamma\gamma)}_{{\bf k},-{\bf k},{\bf 0}}\right)^2 P_{k}^{(\gamma)}P_{k}^{(\gamma)} \nonumber \\
	&=\frac{\pi\epsilon\Delta^2_\zeta}{4}\left(\frac{aH}{q}\right)^3 \ . \label{eq:Gamma_3gamma_sq}
\end{align}
where the product of polarization tensors is evaluated in (\ref{eq:product_polarization_1}) with ${\bf k}\perp {\bf q}$, 
so the decoherence exponent is similar to the scalar case in \cite{Nelson:2016kjm} except different prefactors.

On the other hand, it is possible to have IR divergence when ${\bf k}\to -{\bf q}$ or ${\bf k}'\to -{\bf q}$, leading to terms with $\log\left(\frac{k_{\rm min}}{q}\right)=\Delta N$ regularized by the IR cutoff, and this can be shown from the bulk interaction
\begin{align}
	&\mathcal{L}_{\gamma\gamma\gamma} \nonumber \\
	&\to \frac{M_p^2}{8}a\Bigg\{\lim_{k,k'\to 0}\sum_{s_1,s_2,s_3} 2q_m q_l\left[e^{s_1}_{ij}({\bf k})e^{s_2}_{jm}(-{\bf q})e^{s_3}_{il}({\bf q})\mathcal{E}^{s_1}_{{\bf k}}\mathcal{E}^{s_2}_{-{\bf q}}+e^{s_1}_{il}(-{\bf q})e^{s_2}_{ij}({\bf k}')e^{s_3}_{jm}({\bf q})\mathcal{E}^{s_1}_{-{\bf q}}\mathcal{E}^{s_2}_{{\bf k}'}\right]\xi^{s_3}_{{\bf q}} \nonumber \\
	&+q_iq_j\left[e^{s_1}_{ij}({\bf k})e^{s_2}_{lm}(-{\bf q})\mathcal{E}^{s_1}_{{\bf k}}\mathcal{E}^{s_2}_{-{\bf q}}+e^{s_1}_{lm}(-{\bf q})e^{s_2}_{ij}({\bf k}')\mathcal{E}^{s_1}_{-{\bf q}}\mathcal{E}^{s_2}_{{\bf k}'}\right]	e^{s_3}_{lm}({\bf q})\xi^{s_3}_{{\bf q}} \Bigg\}
	+\mathcal{O}(kk') \nonumber \\
	&\approx M_p^2 a\lim_{k\to 0}\sum_{s,s_3}q_iq_j e^{s}_{ij}({\bf k})\mathcal{E}^{s}_{\bf k}\mathcal{E}^{s_3}_{-{\bf q}} \xi^{s_3}_{{\bf q}}+\mathcal{O}(kk')\ ,
\end{align}
where the factor of polarization tensor is similar to (\ref{eq:squeezed_limit_3gamma}) with ${\bf k} \leftrightarrow {\bf q}$, so it contributes to a factor of $q^4$ to the one-loop integral in this limit. With this, the leading IR logarithmic factor in the decoherence exponent can be estimated as
\begin{align}
	\Gamma^{\rm IR}_{\gamma\gamma\gamma}(q,\tau)&\approx \frac{2H^2M_p^2}{q^3} \int_{k_{\rm min}<k<q} \frac{d^3k}{(2\pi)^3} q^4 \left({\rm Im}I^{(\gamma\gamma\gamma)}_{{\bf k},-{\bf q},{\bf q}}\right)^2 P_{k}^{(\gamma)}P_{q}^{(\gamma)}  \nonumber \\
	&\approx\left(\frac{H}{2M_p\pi}\right)^2\log\left(\frac{q}{k_{\rm min}}\right)\left(\frac{aH}{q}\right)^2 +\mathcal{O}(a^0) \nonumber \\
	&=2\epsilon\Delta^2_\zeta\Delta N \left(\frac{aH}{q}\right)^2 +\mathcal{O}(a^0) \ , \label{eq:Gamma_3gamma_IR}
\end{align}
which agrees with the scalar case \cite{Nelson:2016kjm} that the IR-divergent part is proportional to $a^2$, and adding the dominated part given by the squeezed limit (\ref{eq:Gamma_3gamma_sq}) is the total decoherence exponent
\begin{align}
	\Gamma_{\gamma\gamma\gamma}=\Gamma^{\rm sq}_{\gamma\gamma\gamma}+\Gamma^{\rm IR}_{\gamma\gamma\gamma} \ , \label{eq:Gamma_3gamma}
\end{align} which is sub-dominated compared to the those by boundary terms, as expected in Table \ref{tab:int_terms}.

\subsection{Compare the decoherence by different interaction terms}\label{sec:compare_deco}
We are ready to compare the decoherence by different interactions terms, including those by bulk interactions studied in \cite{Burgess:2022nwu} with $\gamma_{ij}$ and scalar environment, and \cite{Gong:2019yyz} with $h^{TT}_{ij}$ and tensor environment. Since these papers use different quantities to indicate the decoherence, we first convert these to the equivalent decoherence exponent $\Gamma(q,\tau)$ for doing comparison.

For a Gaussian mixed state, the reduced density matrix has the form
\begin{align}
	\rho_R(v,\tilde{v})&=\sqrt{\frac{{\rm Re}A}{\pi}}\exp\left[-\frac{A}{2}v^2-\frac{A^*}{2}\tilde{v}^2-\frac{F^2}{2}\left(v-\tilde{v}\right)^2\right] \ ,
\end{align}
which has the purity 
\begin{align}
	{\rm Tr}\rho_R^2&=\sqrt{\frac{1}{1+\Xi}} \ ,
\end{align}
where $\Xi=\frac{2F^2}{{\rm Re}A}$. In \cite{Burgess:2022nwu}, the decoherence is determined as $\Xi=\mathcal{O}(1)$, and the relation to the decoherence exponent is
\begin{align}
	\Gamma&=\frac{F^2}{2}\left\langle \left(v-\tilde{v}\right)^2 \right\rangle \nonumber \\
	&=\frac{1}{4}\Xi \ .
\end{align}
For the $\zeta\zeta\gamma$ bulk interaction, the decoherence of tensor perturbation is determined by \cite{Burgess:2022nwu}
\begin{align}
	\Xi^{\rm bulk}_{\zeta\zeta\gamma}(q,\tau)&=\frac{1}{36\pi}\left(\frac{H}{M_p}\right)^2\left(\frac{aH}{q}\right)^3 \ ,
\end{align}
which is equivalent to 
\begin{align}
	\Gamma^{\rm bulk}_{\zeta\zeta\gamma}(q,\tau)&=\frac{\pi\Delta^2_\zeta\epsilon}{18}\left(\frac{aH}{q}\right)^3 \ . \label{eq:Gamma_bulk_zeta_zeta_gamma}
\end{align}

On the other hand, \cite{Gong:2019yyz} studies the reduced density matrix $\rho_R$ in the particle basis of the system $\xi$, $\left\{U_{0,\xi}|0\rangle_\xi,U_{0,\xi}a^\dagger_{1,\xi}|0\rangle_\xi,U_{0,\xi}a^\dagger_{1,\xi}a^\dagger_{2,\xi}|0\rangle_\xi\right\}$ where $U_{0,\xi}$ and $a^{\dagger}_{i,\xi}$ are the free evolution and creation operators of the system defined with $h^{\rm TT}_{ij}$, and the evolution of reduced density matrix $\rho_R$ has the form
\begin{align}
	\frac{d\rho_{R}}{d\tau}&=-\frac{1}{2}\begin{pmatrix}
		\mathfrak{E}_{00} & 0 & \mathfrak{E}_{02} \\
		0 & \mathfrak{E}_{11} & 0 \\
		\mathfrak{E}_{20} & 0 & 0 
	\end{pmatrix} + h.c. \ . \label{eq:evolution_rhored_hhh}
\end{align}
The decoherence exponent of the mode ${\bf q}$ illustrated in Fig. \ref{fig:diagrams_gamma_deco}, is comparable to the one with two external 1-particle states and tracing out the sub-horizon tensor environment, which is $\mathfrak{E}_{11}$ in (\ref{eq:evolution_rhored_hhh})\footnote{The authors in \cite{Gong:2019yyz} also comment about this comparison. We should also note that the authors use $\rho_{{\rm red},00}$, calculated with $\mathfrak{E}_{00}$, to indicate the decoherence, but it is obtained by summing over all system modes $\{\xi_{\bf q}\}$ and has IR divergence (as their system is defined as all super-horizon modes), thus not representing a single mode with comoving momentum ${\bf q}$.}
\begin{align}
	\Gamma^{\rm bulk}_{hhh}(q,\tau)&=-\sum_{\lambda,\lambda'=+,\times}\int^\tau_{\tau_i} d\tau' \frac{\mathfrak{E}_{11}(\tau')}{V} \delta_{\lambda,\lambda'}\nonumber \\
	&=\frac{8 H^2}{\pi ^2 M_p^2}\left(\frac{aH}{q}\right)^3 \left\{-\frac{16 }{525 q^3 \tau ^3}\left[-8 q^5 \tau ^5+70 q \tau +35 \log \left(\frac{1-q \tau }{1+q \tau }\right)\right]\right\} \nonumber \\
	&\approx \frac{2048\Delta^2_\zeta\epsilon}{45}\left(\frac{aH}{q}\right)^3 +\mathcal{O}\left(a^2\right) \ .  \label{eq:Gamma_bulk_hhh}
\end{align}

\begin{figure}[h!]
	\centering
	\includegraphics[width=0.7\textwidth]{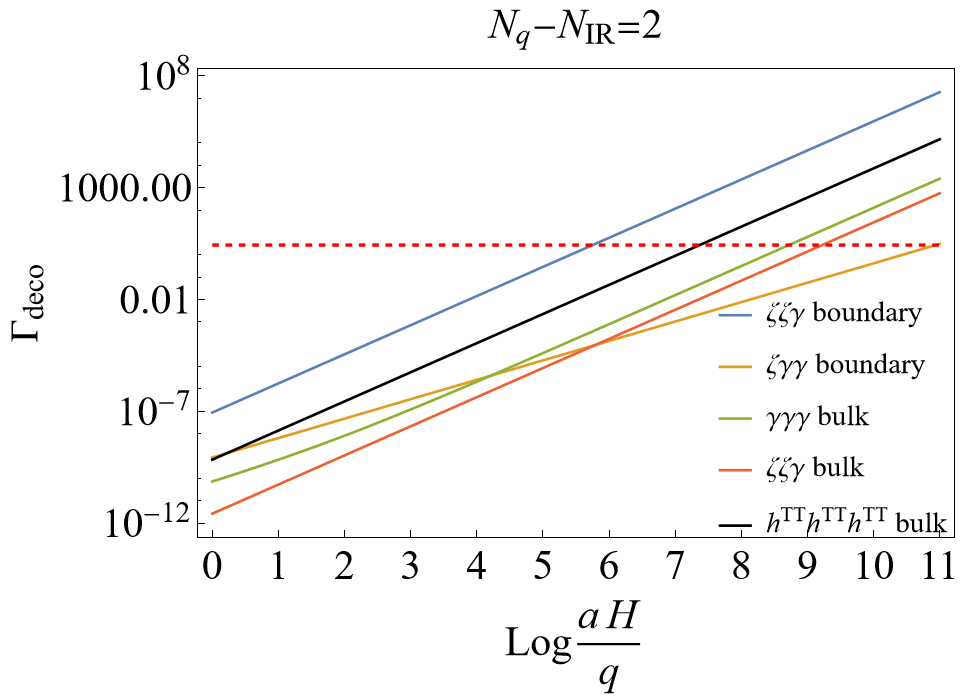}
	\includegraphics[width=0.7\textwidth]{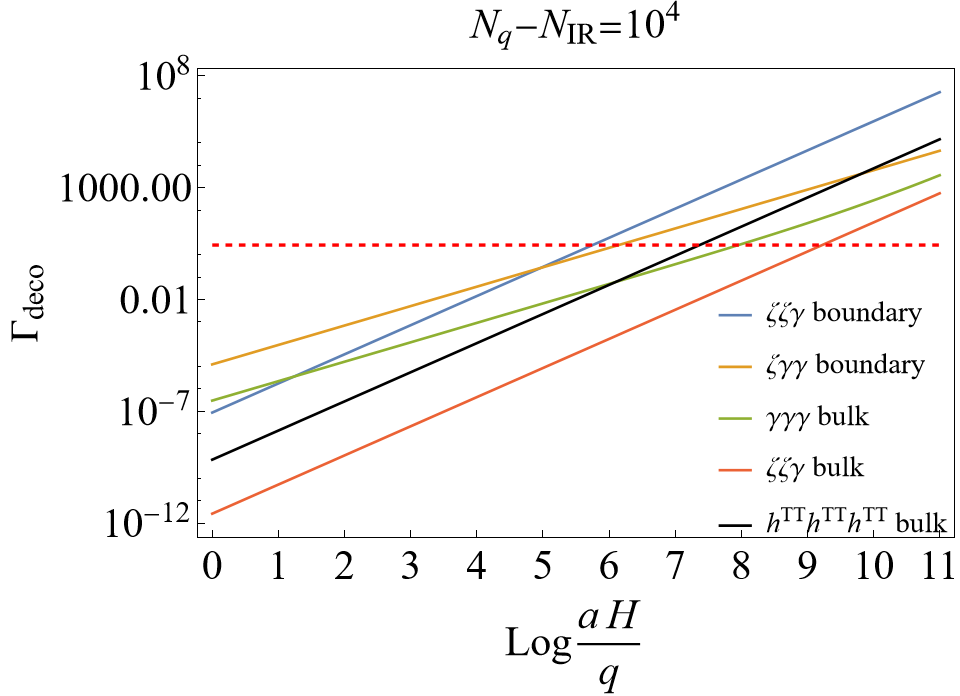}
	\caption{The comparison of $\Gamma(q,\tau)$ from different interactions (\ref{eq:Gamma_zeta_zeta_gamma_renormalized}), (\ref{eq:Gamma_zeta_gamma_gamma_renormalized}), (\ref{eq:Gamma_3gamma}), (\ref{eq:Gamma_bulk_zeta_zeta_gamma}) and (\ref{eq:Gamma_bulk_hhh}). The red-dashed line indicates $\Gamma(q,\tau)=1$, and the observed values $\Delta_\zeta^2\approx2.5\times 10^{-9}$ and $\epsilon<0.006$ are used \cite{Planck:2018jri}. Two values of the IR cutoff are chosen: $\Delta N=2$ for the minimal inclusion of the IR environment, and  $\Delta N=10^4$ as the value used in \cite{Martin:2018zbe}.} \label{fig:comparision_5_Gamma_graviton}
\end{figure}

Fig. \ref{fig:comparision_5_Gamma_graviton} shows the comparison of $\Gamma(q,\tau)$ (\ref{eq:Gamma_zeta_zeta_gamma_renormalized}), (\ref{eq:Gamma_zeta_gamma_gamma_renormalized}), (\ref{eq:Gamma_3gamma}), (\ref{eq:Gamma_bulk_zeta_zeta_gamma}) and (\ref{eq:Gamma_bulk_hhh}) from different interaction terms, and two values of the IR cutoff $\log\left(\frac{q}{k_{\rm min}}\right)=\Delta N$ are chosen: $\Delta N=2$ for the minimal value for making the valid expansion with small $\frac{k_{\rm min}}{q}$ and $\Delta N=10^4$ as the value used in \cite{Martin:2018zbe}. For the decoherence happens with $\Gamma(q,\tau)\approx 1$, the cases with bulk interactions require 7-9 e-folds after the horizon crossing, where such a 1-2 e-fold difference is partly attributed to different number prefactors obtained by various methods of calculating decoherence,\footnote{Note that (\ref{eq:Gamma_3gamma_sq}), (\ref{eq:Gamma_bulk_zeta_zeta_gamma}) and (\ref{eq:Gamma_bulk_hhh}) have the same order of slow-roll suppression.} and it is also attributed to the IR part (\ref{eq:Gamma_3gamma_IR}) by the bulk interaction $\mathcal{L}_{\gamma\gamma\gamma}$. On the other hand, the decoherence by the boundary terms happens around 5-6 e-folds after the horizon crossing, which is faster than the cases with the bulk interactions, expected by counting the order of slow-roll suppression in Table \ref{tab:int_terms}.

\section{Comments on the one-loop quantum correction} \label{sec:1_loop_correction}
Previously, our treatment is limited to the tree-level analysis, and it remains to be seen how quantum fluctuation will influence the result. Since gravity can be viewed as a non-abelian gauge theory with diffeomorphism invariance, we need to also include ghost fields. In this section, we give an argument that the effect of one-loop correction is neglectable compared to the leading order contribution.

To begin with, following (\ref{eq:action_single_field}), the corresponding Euclidean action reads\footnote{In the following calculation of effective action, we follow the metric signature in \cite{tHooft:1974toh}.}
\begin{align}
	&I[g,\phi]=\int d^4x \ \mathcal{L} \nonumber \\
	&=\int_{\mathcal{M}}d^4x \sqrt{g}\left[-\frac{M_p^2}{2}R-\frac{1}{2}g^{\mu\nu}\partial_\mu\phi\partial_\nu\phi+V(\phi)\right]-\int_{\partial \mathcal{M}}d^3x M_p^2\sqrt{h}K \ ,\quad \mu,\nu=1,2,3,4 \label{Eucldean1}
\end{align}  
and the wave functional can be expressed with the path integral
\begin{equation}
	\Psi(h,\phi_0)=\int_{g|_{\partial \mathcal{M}}=h, \ \phi|_{\partial \mathcal{M}}=\phi_0}\mathcal{D}g \mathcal{D}\phi \ e^{- I[g,\phi]} \ ,
\end{equation}   
where $h$ and $\phi_0$ simply denote the boundary values of the metric and inflation field. The metric and inflation field can be expanded as
\begin{equation}
	g=g_c+ \tilde{h} \ ,\quad \phi=\phi_c+\varphi \ ,
\end{equation}
where $g_c$ and $\phi_c$ are the solutions to the classical equations of motion
\begin{equation}
	\frac{\delta I[g,\phi]}{\delta g}=0 \ , \quad \frac{\delta I[g,\phi]}{\delta \phi}=0 \ ,
\end{equation}
and under the diffeomorphism $x_\mu\rightarrow x_\mu+\xi_\mu$ the perturbations transform as
\begin{equation}
	\tilde{h}_{\mu\nu}\rightarrow \tilde{h}_{\mu\nu}+\nabla_\mu\xi_\nu+\nabla_\nu \xi_\mu \ ,\quad \varphi\rightarrow \varphi-\xi^\mu\nabla_\mu\varphi \ .
\end{equation}
Following \cite{Witten:2018lgb, Witten:2022xxp}, the boundary condition here is subtle. Here we mainly focus on metric field because it is the quantum gravity that cause the subtleties instead of scalar field. Usually there are two boundary conditions for metric field, one is the Dirichlet boundary condition, which fixes the metric at the boundary and requires $\tilde{h}|_{\partial \mathcal{M}}=0$, $\xi|_{\partial \mathcal{M}}=0$, so the transverse component of metric is unrestricted and needs gauge fixing. The other is called conformal boundary condition which admits the boundary metric up to a Weyl rescaling. In \cite{Witten:2018lgb}, the author points out that mathematically, the Dirichlet boundary condition will be problematic because this boundary condition breaks the elliptic properties of propagator, and therefore, may not lead to a well-defined perturbation theory of quantum gravity. The conformal boundary condition will satisfy the elliptic properties. However, the author also said that, a sufficient condition for the Dirichlet boundary continues to work is that the extrinsic curvature $K_{ij}$ is either positive or negative definite. For de Sitter it is not hard to prove that the metric satisfies this condition, and therefore both the Dirichlet and conformal boundary condition will work for the analysis.

Without loss of generality, we take the Dirichlet boundary condition, and we need to fix the transverse component of the metric. Following \cite{tHooft:1974toh}, the gauge condition reads:
\begin{equation}
	C_{\alpha}={g_c}^{\frac{1}{4}}\left(\nabla_{\nu} \tilde{h}^{\nu}_{\mu}-\frac{1}{2}\nabla_{\mu} \tilde{h}-\varphi\nabla_{\mu}\phi_c\right)t^{\mu}_{\alpha}=0, \quad \alpha=1,2,3,4\label{Faddev1}
\end{equation}
where $t^{\mu}_{\alpha}t^{\alpha \nu}=g_c^{\mu\nu}$. It is worthy to note that some references \cite{FRADKIN1984472,Bamba:2014mua} also take the Landau gauge. For the on-shell effective action, different choices of $C_\alpha$ are equivalent, whereas for the off-shell effective action, the one-loop correction depends on the choice of gauge. However, we will show that in our case, higher loop corrections can be neglected, and therefore it is reasonable to only consider the on-shell effective action, in which the choice of gauge will not influence our conclusion.

By expanding (\ref{Faddev1}) around an infinitesimal gauge transformation, following \cite{tHooft:1974toh}, the ghost Lagrangian density for one loop is given as:
\begin{equation}
	\mathcal{L}_{\rm ghost}=\sqrt{g_c}\bar{c}^{\mu}\left(\nabla_a\nabla^a\delta_{\mu}^{\nu}-R_{\mu}^{\nu}-\nabla^{\nu}\phi_c\nabla_{\mu}\phi_c\right)c_v=\sqrt{g_c}\bar{c}^{\mu} J_{\mu}^{\nu} c_{\nu} \ , \label{eq:L_ghost}
\end{equation}
where $c$ and $\bar{c}$ are ghost and anti-ghost fields respectively, which obey the Dirichlet boundary condition and (\ref{Faddev1}). Then we expand (\ref{Eucldean1}) to the second order of the metric and inflation field and adding the gauge-fixing term, following \cite{tHooft:1974toh}
\begin{equation}
	\begin{aligned}
	 \mathcal{L}_{\rm gf}=\mathcal{L}-\frac{1}{2}C_{\mu} C^{\mu}&=\sqrt{g_c}\Bigg(-\frac{1}{4}\tilde{h}^{\beta}_{\alpha,\nu}\tilde{h}^{\alpha,\nu}_{\beta}+\frac{1}{8}\tilde{h}^{\alpha}_{\alpha,\nu}\tilde{h}^{\beta,\nu}_{\beta}-\frac{1}{2}\nabla_\mu\varphi \nabla^\mu\varphi\\
	&+\frac{1}{2}\tilde{h}_{\beta}^{\alpha}X_{\alpha \nu}^{\beta \mu}\tilde{h}_{\mu}^{\nu}+\varphi Y_{\beta}^{\alpha}\tilde{h}_{\alpha}^{\beta}+\frac{1}{2}\varphi Z\varphi \Bigg) \ ,
	\end{aligned}
\end{equation}
where the definitions of $X,Y,Z$ are given in \cite{tHooft:1974toh}:
\begin{equation}
	\begin{aligned}
		 X_{\alpha \nu}^{\beta \mu}&=2\Bigg\{-\frac{1}{2}\delta_{\nu}^{\beta}\nabla^{\mu}\phi_c \nabla_{\alpha}\phi_c+\frac{1}{4}\delta_{\alpha}^{\beta}\nabla^{\mu}\phi_c\nabla_{\nu}\phi_c-\frac{1}{16}\delta_{\alpha}^{\beta}\delta_{\nu}^{\mu}\left[\nabla_{\gamma}\phi_c\nabla^{\gamma}\phi_c+2V(\phi_c)\right]\\
		&+\frac{1}{8}\delta_{\nu}^{\beta}\delta_{\alpha}^{\mu}\left[\nabla_{\gamma}\phi_c\nabla^{\gamma}\phi_c+2V(\phi_c)\right]-\frac{1}{8}\delta_{\alpha}^{\beta}\delta_{\nu}^{\mu}R+\frac{1}{4}\delta^{\beta}_{\nu}\delta_{\alpha}^{\mu} R\\
		&-\frac{1}{2}\delta_{\nu}^{\beta}R_{\alpha}^{\mu}+\frac{1}{2}\delta_{\alpha}^{\beta}R_{\nu}^{\mu}+\frac{1}{2}R^{\beta \mu}_{\quad\alpha \nu}\Bigg\},\\
		 Y_{\beta}^{\alpha}&=\frac{1}{2}\delta_{\beta}^{\alpha}\nabla_{\nu} \nabla^{\nu}\phi_c- \nabla_{\beta}\nabla^{\alpha}\phi_c+\frac{1}{2}\frac{\delta V}{\delta \phi}\Big|_{\phi=\phi_c}\delta^{\alpha}_{\beta},\\
		 Z&=-\nabla_{\mu}\phi_c\nabla^{\mu}\phi_c+\frac{1}{2}\frac{\partial^2 V}{\partial\phi^2}\Big|_{\phi=\phi_c} \ ,
	\end{aligned}
\end{equation}
and we used the facts that:
\begin{equation}
	\delta\sqrt{g}=\frac{1}{2}\sqrt{g_c}\tilde{h}_{\alpha}^{\alpha}+\frac{1}{4}\sqrt{g_c}\left[-\tilde{h}^{\alpha}_{\beta}\tilde{h}^{\beta}_{\alpha}+\frac{1}{2}(\tilde{h}^{\alpha}_{\alpha})^2\right] \ ,
\end{equation}
and
\begin{equation}
	\delta V(\phi)=\frac{\partial V}{\partial \phi}\Big|_{\phi=\phi_c}\varphi+\frac{1}{2}\frac{\partial^2 V}{\partial\phi^2}\Big|_{\phi=\phi_c}\varphi^2 \ .
\end{equation}
Following \cite{Barvinsky:2006gd,Freidel:2008sh}, the one-loop wave functional is then formally written as:
\begin{equation}
	\Psi_{{\rm1-loop}}=\frac{\text{det}{\bf J}}{(\text{det}{\bf G})^{\frac{1}{2}}}e^{-I[\phi_c, g_c]}=e^{-I_{\rm eff}[\phi_c, g_c]} \ ,
\end{equation}
where ${\bf J}$ is defined in (\ref{eq:L_ghost}), and ${\bf G}$ is defined as:
\begin{equation}
	{\bf G}=\begin{pmatrix}
		\frac{\delta^2 \mathcal{L}_{\rm gf}}{\delta^2 \tilde{h}_{\alpha\beta}\tilde{h}_{\mu\nu}}&&\frac{\delta^2 \mathcal{L}_{\rm gf}}{\delta \tilde{h}_{\mu\nu}\delta \varphi}\\
		\frac{\delta^2 \mathcal{L}_{\rm gf}}{\delta \tilde{h}_{\alpha\beta}\delta \varphi}&&\frac{\delta^2 \mathcal{L}_{\rm gf}}{\delta\varphi^2} 
	\end{pmatrix} \ ,
\end{equation}
and the effective action is written as:
\begin{equation}
	I_{\rm eff}[\phi_c, g_c]=I[\phi_c, g_c]-\text{Tr} \log{\bf J}[\phi_c, g_c]+\frac{1}{2}\text{Tr} \log {\bf G}[\phi_c, g_c] \ .
\end{equation}

The explicit calculation of the determinants needs to use the zeta function regulator and is model dependent, because here we need to know the explicit form of inflaton potential $V(\phi)$. The results have been shown in various literature, for reference, see \cite{Gibbons:1976ue,Gibbons:1978ac, Christensen:1979iy, FRADKIN1984472,Bamba:2014mua}. Here we take the example in \cite{FRADKIN1984472}: the effective action for a pure gravity with de Sitter background has the form
\begin{equation}
	I_{\rm eff}\sim \int d^4x\sqrt{g}\left[-\frac{M_p^2}{2}(R-2\Lambda)+R^2\log\left(\frac{R}{\mu^2}\right)\right] \ , \label{eq:form_of_1_loop_Gamma}
\end{equation}
where the first term is tree-level, and the second term is the one-loop correction with a energy scale $\mu$. For the de Sitter space with $R=4\Lambda=12H^2$, the ratio of the one-loop to tree-level contributions is:
\begin{align}
	\frac{I_{\rm 1-loop}}{I_{\rm tree}}&\sim \frac{H^2}{M_p^2}\log\frac{R}{\mu^2} \sim \Delta^2_\gamma \lesssim \mathcal{O}\left(10^{-10}\right) \ , \label{eq:1loop_order_estimation}
\end{align}
where the energy scale is chosen as $\mu\sim H$, and the Planck data \cite{Planck:2018jri} is used for the estimation. Similar form of the one-loop correction (\ref{eq:form_of_1_loop_Gamma}) is also derived in explicit inflation models such as \cite{Bamba:2014mua}, so the order estimation (\ref{eq:1loop_order_estimation}) should be generic. Such a small ratio implies that the one-loop and higher-order corrections to the phase of wave functional and the corresponding decoherence effect are neglectable.

Following \cite{Weinberg:2005vy,Senatore:2009cf,Pimentel:2012tw,Senatore:2012nq}, it is worthy to mention that the symmetry also allows some other one-loop corrections, take the two-point function of scalar curvature perturbation as an example, there are two possible terms that allowed by symmetry. One is:
\begin{equation}
    \langle\zeta_p^2\rangle_{\rm 1-loop}\sim p^3\langle\zeta_{p}^2\rangle_{\rm tree}^2\log(pL) \ ,
\end{equation}
where $L$ being the comoving size, and the other is:
\begin{equation}
    \langle\zeta_p^2\rangle_{\rm 1-loop}\sim p^3\langle\zeta_{p}^2\rangle_{\rm tree}^2\log(a(t)) \ .
\end{equation}
The first term might contribute large in the IR limit, however, \cite{Senatore:2012nq} shows that this will not affect the observable quantities and is purely a projection effect. The second term is only logarithmic type with the suppression of $\Delta^2_\zeta$ that it will not influence the result significantly as well. Similarly, the one-loop correction of gravitons' two-point function also includes a logarithmic term like \cite{Brahma:2022yxu}
\begin{equation}
    \left\langle \left(h^{\rm TT}_{p}\right)^2\right\rangle_{\rm 1-loop}\sim p^3\left \langle \left(h^{\rm TT}_{p}\right)^2\right\rangle_{\rm tree}^2\log(a(t)) \ ,
\end{equation}
which also does not correct significantly to the result, following the reason of the scalar case.

\section{Conclusion} \label{sec:conclusion}
Studying the decoherence of primordial gravitons is not only for explaining the quantum-to-classical transition of the primordial gravitational wave, but it is also useful to set constraints and identify potential obstacles for probing the non-classicality of squeezed gravitons. In this paper, we proceeded the discussion of the slow-roll unsuppressed boundary-term decoherence in \cite{Sou:2022nsd} and calculated the case with primordial gravitons. 

Starting from the standard procedure of splitting the bulk and boundary interaction terms with integration by parts, we confirmed that there exists unsuppressed scalar-tensor coupling boundary terms of $\zeta$ and $\gamma_{ij}$. Such boundary terms were shown to contribute a non-Gaussian phase to the wave functional of cosmological perturbations with either the interaction and Schr\"odinger picture approaches. As the phase grows with the scale factor $a(t)$, the corresponding quantum state is close to the WKB type, as expected in \cite{Nelson:2016kjm}, but now we have showed that it is slow-roll unsuppressed.

To gain insight into the unsuppressed phase of the wave functional, we studied the Wheeler-DeWitt formalism of quantum gravity with the WKB approximation valid in the large $a(t)$ limit. The WKB phase evaluated on the hypersurface with the given $(h_{ij},\phi)$ was shown to include unsuppressed cubic terms, agreeing with those obtained from the boundary terms. This result suggests that the leading phase of the inflationary wave functional is independent to the bulk-boundary splitting of the action, but such a splitting is certainly helpful to identify the unsuppressed terms. The agreement of these approaches also suggests that full considerations of either the boundary terms of cosmological perturbations or the WKB limit of the WDW wave functional are needed to discuss the classicalization of the perturbations. 

We thus calculated the environment-induced decoherence of primordial gravitons with the wave functional method presented in \cite{Nelson:2016kjm}, addressing the influence by the unsuppressed non-Gaussian phase. The gravitons were shown to decohere around 5-6 e-folds after crossing the horizon, faster than the process by the bulk interactions which takes 7-9 e-folds. We have also estimated the size of the one-loop quantum correction to the wave functional including the Faddeev-Popov ghost determinant, showing that it is suppressed by a factor of graviton's power spectrum $\frac{H^2}{M_p^2}$ compared to the three-level part, so the correction to our results should be negligible. During inflation, the squeezing of primordial gravitons happens after the horizon crossing with the parameter grows as $R_q\sim\log\left(\frac{aH}{q}\right)$ \cite{Albrecht:1992kf}, so their purity should also be considered in the proposals testing their non-classicality. How exactly the cosmological decoherence affects the proposals deserves further studies.

Finally, we comment on the relationship between the physical hypersurface $\Sigma$ and decoherence. As demonstrated in \cite{Sou:2022nsd} with the well-defined variational principle, the boundary terms on the two hypersurfaces naturally defined in the $\zeta$-gauge and $\delta\phi$-gauge respectively are different, whereas they become equal when the same $\Sigma$ is chosen even if the calculation is done with the fields defined in different gauges. It is noteworthy that similar results have been reported in \cite{Prokopec:2013zya} by constructing gauge-invariant quantities for all the scalar and tensor interactions, and boundary terms on different hypersurfaces do not match even the non-linear field redefinition is considered. Here we make another point of view by studying the phase of the WDW wave functional, and the leading WKB phase $W(h_{ij},\phi)$ is evaluated by choosing $\Sigma$. By the relationship between the phase of wave functional and boundary terms, it is easy to verify the facts in \cite{Prokopec:2013zya,Sou:2022nsd}. This also implies the same conclusion in \cite{Sou:2022nsd} that the quantities naturally defined on various $\Sigma$ have different decoherence rates, such as $\zeta$ and $\gamma_{ij}$ which conserve on super-horizon scales decohere faster. Better understanding the physical meanings of these conclusions is also worthy for further studies. Following \cite{Maldacena:2002vr,DelCampo:2019afl}, it will also be interesting to understand the decoherence effect in this paper from the perspective of dS/CFT and AdS/CFT, and in the context of string cosmology \cite{Cicoli:2023opf}. We leave this for future work.

\section*{Acknowledgments}
We thank Ali Akil, Haipeng An, Chao Chen, Joseph Conlon, Ling-Yan Hung, Juan Maldacena, Rok Medves, Michael Nee, Malcolm Perry, Guilherme L. Pimentel, Xi Tong, Aron Wall, Zhenbin Yang, Zhong-Zhi Xianyu and Yuhang Zhu for helpful discussions. SN wants to acknowledge funding support from the China Scholarship Council-FaZheng Group- University of Oxford. CMS and YW were supported in part by the National Key R\&D Program of China (2021YFC2203100), the NSFC Excellent Young Scientist Scheme (Hong Kong and Macau) Grant No. 12022516, and by the RGC of Hong Kong SAR, China
(Project No. 16306422 and 16303621).

\newpage

\appendix
\section{The analytical expressions of $J_{\zeta\zeta\gamma}(Q)$ and $J_{\zeta\gamma\gamma}(Q)$}
\begin{figure}
	\centering
	\includegraphics[width=0.47\textwidth]{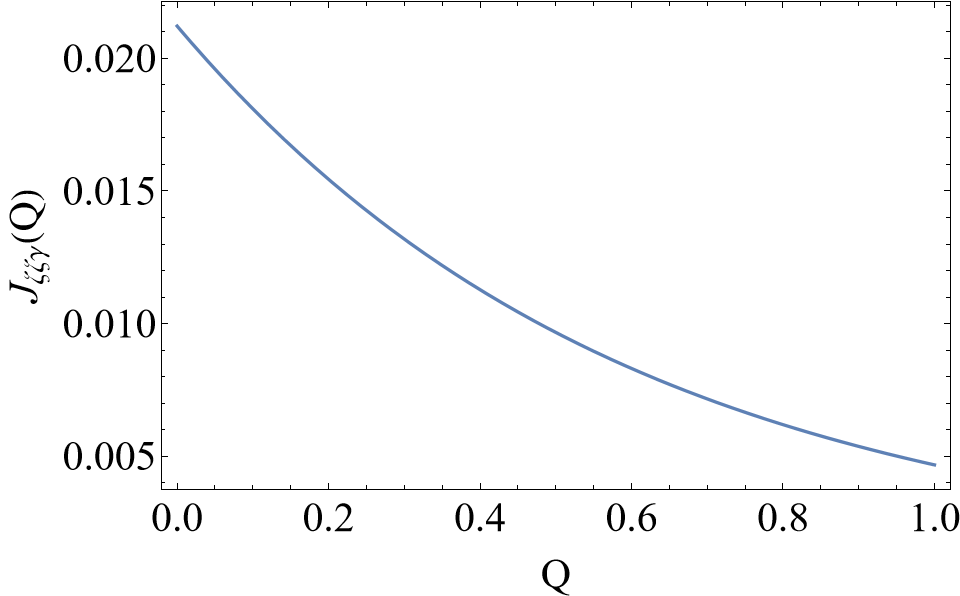}
	\includegraphics[width=0.47\textwidth]{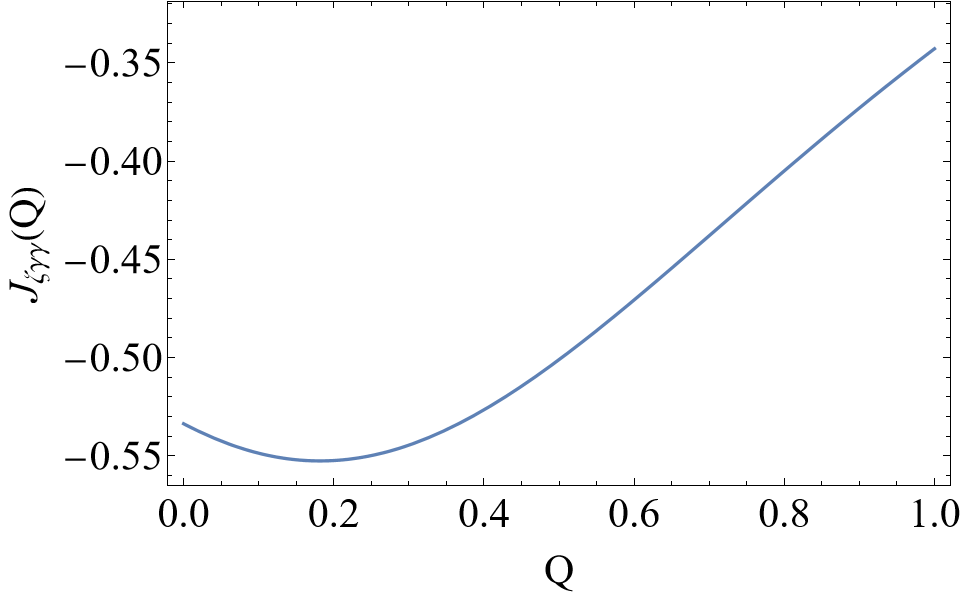}
	\caption{$J_{\zeta\zeta\gamma}\left(Q\right)$ and $J_{\zeta\gamma\gamma}\left(Q\right)$ defined in (\ref{eq:Gamma_zeta_zeta_gamma_renormalized}) and (\ref{eq:Gamma_zeta_gamma_gamma_renormalized}) respectively.}\label{fig:J_zeta_zeta_gamma_and_zeta_gamma_gamma}
\end{figure}

The expression of $J_{\zeta\zeta\gamma}$ in (\ref{eq:Gamma_zeta_zeta_gamma_renormalized}) is
\begin{align}
		&J_{\zeta\zeta\gamma}(Q)=-\frac{1}{3840 \pi ^2 Q^5}\Bigg\{8 \left(15 Q^6+90 Q^4+80 Q^2+48\right) \log \left(Q^2+1\right) \nonumber \\
		&-8 Q^2 \left[15 \pi  Q^5-30 Q^4-28 \pi  Q^3+56 Q^2+\left(56 Q^3-30 Q^5\right)\tan ^{-1}(Q)+48\right] \nonumber \\
		&+15 Q^4 \left(Q^2+4\right)^2 \Bigg[\text{Li}_2\left(\frac{Q (Q-i)}{Q^2-i Q+2}\right)+\text{Li}_2\left(\frac{Q (Q+i)}{Q^2+i
			Q+2}\right)-\text{Li}_2\left(\frac{Q^2+i Q+2}{Q^2+i Q}\right)\nonumber \\
		&+2 \text{Li}_2\left(\frac{2 i}{Q}-1\right)-\text{Li}_2\left(\frac{Q}{Q-2
			i}\right)-\text{Li}_2\left(\frac{Q-2 i}{Q}\right)-\text{Li}_2\left(\frac{(Q+i) (Q-2 i)}{Q (Q-i)}\right) \nonumber \\
		&-2 \text{Li}_2\left(-\frac{Q}{Q+2
			i}\right)+\text{Li}_2\left(\frac{Q}{Q+2 i}\right)+\text{Li}_2\left(\frac{Q+2 i}{Q}\right)+2 i \log \left(\frac{2}{Q^2-i Q+2}\right) \tan
		^{-1}(Q) \nonumber \\
		&-2 i \log \left(-\frac{2}{Q^2+i Q}\right) \tan ^{-1}(Q)-2 i \log \left(\frac{2}{Q^2+i Q+2}\right) \tan ^{-1}(Q)+i \pi  \log \left(-\frac{2 i}{Q}\right) \nonumber \\
		&-i \pi  \log \left(\frac{2 i}{Q}\right)-i \pi  \log
		\left(-\frac{2 i}{Q-2 i}\right)+i \pi  \log \left(\frac{2 i}{Q+2 i}\right)+2 i \log \left(-\frac{2}{Q (Q-i)}\right) \tan ^{-1}(Q) \nonumber \\
		&+\mathcal{T}_{\zeta\zeta\gamma}^+(Q) \tanh ^{-1}(1+i Q)+\mathcal{T}_{\zeta\zeta\gamma}^-(Q)\tanh
		^{-1}(1-i Q)\Bigg]
		\Bigg\} \ , \label{eq:explicit_J_zeta_zeta_gamma}
\end{align} 
where 
\begin{align}
	\mathcal{T}_{\zeta\zeta\gamma}^+(Q)&=-\log \left(4 Q (Q-2 i) \left(Q^2+1\right)\right)+2 \log \left(-\frac{i}{\sqrt{Q (Q-2 i) \left(Q^2+1\right)}}\right)+2 \log (-1+i Q) \nonumber \\
	&+2 \log
	(2+i Q)+4 \log (2+2 i Q)+2 \log (Q-i)+2 \log \left(\frac{i}{2 (Q-2 i)}\right) \nonumber \\
	&-4 \log \left(\frac{Q-i}{\sqrt{Q (Q-2 i)}}\right)-4 i\pi \ , \nonumber \\
	\mathcal{T}_{\zeta\zeta\gamma}^-(Q)&=4 \log (1-i Q)+2 \log (Q+i)-2 \log \left(\frac{2 i}{Q+2 i}\right)-4 \log \left(\frac{Q+i}{\sqrt{Q+2 i}}\right) \nonumber \\
	&+2 \log \left(\frac{i}{2\sqrt{Q+i} \sqrt{2+Q (Q+i)}}\right)+2 \log \left(-\frac{4 i}{\sqrt{Q+i} \sqrt{2+Q (Q+i)}}\right) \nonumber \\
	&-4 \tanh ^{-1}\left(\frac{1}{3}-\frac{2 i
		Q}{3}\right)-3 i \pi \ .
\end{align}
On the other hand, $J_{\zeta\gamma\gamma}(Q)$ defined in (\ref{eq:Gamma_zeta_gamma_gamma_renormalized}) is
\begin{align}
	&J_{\zeta\gamma\gamma}(Q)
	=-\frac{1}{1935360 \pi ^2 Q^6 \left(Q^2+1\right)} \Bigg\{ \nonumber \\
	&24 \left(315 Q^{12}-7875 Q^{10}+13650 Q^8-23835 Q^6+82296 Q^4-32491 Q^2+818\right) \log \left(Q^2+1\right) \nonumber \\
	&-16 \Bigg[9984+43008 Q^2+\Bigg(3 \pi  Q \left(315 Q^6-10225 Q^4+44756 Q^2-25344\right) \nonumber \\
	&-2 \left(945 Q^6-30675 Q^4+135988 Q^2+151096\right)\Bigg)
	Q^4\Bigg] Q^2 \nonumber \\
	&+\mathcal{T}_{\zeta\gamma\gamma}(Q) \log (1+i Q)+\mathcal{T}^*_{\zeta\gamma\gamma}(Q)
	\log (1-i Q)+945 Q^8 \left(Q^6-23 Q^4-8 Q^2+16\right) \Bigg[4 \text{Li}_2\left(-\frac{i}{Q}\right) \nonumber \\
	&+4 \text{Li}_2\left(\frac{i}{Q}\right)-4
	\text{Li}_2\left(\frac{Q-i}{Q}\right)-4 \text{Li}_2\left(\frac{Q+i}{Q}\right)-4 \text{Li}_2\left(-\frac{i}{Q-2 i}\right)+4
	\text{Li}_2\left(\frac{Q-i}{Q-2 i}\right) \nonumber \\
	&-4 \text{Li}_2\left(\frac{i}{Q+2 i}\right)+4 \text{Li}_2\left(\frac{Q+i}{Q+2 i}\right)+\log
	^2(-Q-2 i)+\log ^2(-Q+2 i)-\log ^2(Q-2 i) \nonumber \\
	&-\log ^2(Q+2 i)-i \pi  \log (-Q-2 i)+i \pi  \log (-Q+2 i)+i \pi  \log (Q-2 i)-i \pi  \log (Q+2
	i)\Bigg]
	\Bigg\} \ , \label{eq:explicit_J_zeta_gamma_gamma}
\end{align}
where
\begin{align}
	\mathcal{T}_{\zeta\gamma\gamma}(Q)&=6 \Bigg[-315 i \pi  Q^{14}-2520 i Q^{13}+(1260+7245 i \pi ) Q^{12}+81800 i Q^{11} \nonumber \\
	&+(-31500+2520 i \pi ) Q^{10}-358048 i Q^9+(54600-5040 i \pi
	) Q^8 +202752 i Q^7 \nonumber \\
	&+471660 Q^6+618912 Q^4+257964 Q^2+23352 \nonumber \\
	&+630 \left(Q^6-23 Q^4-8 Q^2+16\right) Q^8 \log (Q) \nonumber \\
	&+630 \left(Q^6-23 Q^4-8 Q^2+16\right)
	Q^8 \log \left(-\frac{i}{Q-2 i}\right)\Bigg] \ .
\end{align}
The two functions have real values on $0<Q=\frac{q}{aH}<1$, shown in Fig \ref{fig:J_zeta_zeta_gamma_and_zeta_gamma_gamma}.

\bibliographystyle{utphys}
\bibliography{reference}

\providecommand{\href}[2]{#2}\begingroup\raggedright\begin{thebibliography}{10}

\bibitem{Grishchuk:1989ss}
L.~P. Grishchuk and Y.~V. Sidorov, ``{On the Quantum State of Relic
  Gravitons},'' \href{http://dx.doi.org/10.1088/0264-9381/6/9/002}{{\em Class.
  Quant. Grav.} {\bfseries 6} (1989) L161--L165}.

\bibitem{Grishchuk:1990bj}
L.~P. Grishchuk and Y.~V. Sidorov, ``{Squeezed quantum states of relic
  gravitons and primordial density fluctuations},''
  \href{http://dx.doi.org/10.1103/PhysRevD.42.3413}{{\em Phys. Rev. D}
  {\bfseries 42} (1990) 3413--3421}.

\bibitem{Albrecht:1992kf}
A.~Albrecht, P.~Ferreira, M.~Joyce, and T.~Prokopec, ``{Inflation and squeezed
  quantum states},'' \href{http://dx.doi.org/10.1103/PhysRevD.50.4807}{{\em
  Phys. Rev. D} {\bfseries 50} (1994) 4807--4820},
  \href{http://arxiv.org/abs/astro-ph/9303001}{{\ttfamily
  arXiv:astro-ph/9303001}}.

\bibitem{Polarski:1995jg}
D.~Polarski and A.~A. Starobinsky, ``{Semiclassicality and decoherence of
  cosmological perturbations},''
  \href{http://dx.doi.org/10.1088/0264-9381/13/3/006}{{\em Class. Quant. Grav.}
  {\bfseries 13} (1996) 377--392},
  \href{http://arxiv.org/abs/gr-qc/9504030}{{\ttfamily arXiv:gr-qc/9504030}}.

\bibitem{Parikh:2020nrd}
M.~Parikh, F.~Wilczek, and G.~Zahariade, ``{The Noise of Gravitons},''
  \href{http://dx.doi.org/10.1142/S0218271820420018}{{\em Int. J. Mod. Phys. D}
  {\bfseries 29} no.~14, (2020) 2042001},
  \href{http://arxiv.org/abs/2005.07211}{{\ttfamily arXiv:2005.07211
  [hep-th]}}.

\bibitem{Parikh:2020fhy}
M.~Parikh, F.~Wilczek, and G.~Zahariade, ``{Signatures of the quantization of
  gravity at gravitational wave detectors},''
  \href{http://dx.doi.org/10.1103/PhysRevD.104.046021}{{\em Phys. Rev. D}
  {\bfseries 104} no.~4, (2021) 046021},
  \href{http://arxiv.org/abs/2010.08208}{{\ttfamily arXiv:2010.08208
  [hep-th]}}.

\bibitem{Parikh:2020kfh}
M.~Parikh, F.~Wilczek, and G.~Zahariade, ``{Quantum Mechanics of Gravitational
  Waves},'' \href{http://dx.doi.org/10.1103/PhysRevLett.127.081602}{{\em Phys.
  Rev. Lett.} {\bfseries 127} no.~8, (2021) 081602},
  \href{http://arxiv.org/abs/2010.08205}{{\ttfamily arXiv:2010.08205
  [hep-th]}}.

\bibitem{Kanno:2020usf}
S.~Kanno, J.~Soda, and J.~Tokuda, ``{Noise and decoherence induced by
  gravitons},'' \href{http://dx.doi.org/10.1103/PhysRevD.103.044017}{{\em Phys.
  Rev. D} {\bfseries 103} no.~4, (2021) 044017},
  \href{http://arxiv.org/abs/2007.09838}{{\ttfamily arXiv:2007.09838
  [hep-th]}}.

\bibitem{Haba:2020jqs}
Z.~Haba, ``{State-dependent graviton noise in the equation of geodesic
  deviation},'' \href{http://dx.doi.org/10.1140/epjc/s10052-020-08805-y}{{\em
  Eur. Phys. J. C} {\bfseries 81} no.~1, (2021) 40},
  \href{http://arxiv.org/abs/2009.12306}{{\ttfamily arXiv:2009.12306 [gr-qc]}}.

\bibitem{Cho:2021gvg}
H.-T. Cho and B.-L. Hu, ``{Quantum noise of gravitons and stochastic force on
  geodesic separation},''
  \href{http://dx.doi.org/10.1103/PhysRevD.105.086004}{{\em Phys. Rev. D}
  {\bfseries 105} no.~8, (2022) 086004},
  \href{http://arxiv.org/abs/2112.08174}{{\ttfamily arXiv:2112.08174 [gr-qc]}}.

\bibitem{Bak:2022oyn}
S.-E. Bak, M.~Parikh, S.~Sarkar, and F.~Setti, ``{Quantum Gravity Fluctuations
  in the Timelike Raychaudhuri Equation},''
  \href{http://arxiv.org/abs/2212.14010}{{\ttfamily arXiv:2212.14010 [gr-qc]}}.

\bibitem{Cho:2023dmh}
H.-T. Cho and B.-L. Hu, ``{Graviton noise on tidal forces and geodesic
  congruences},'' \href{http://dx.doi.org/10.1103/PhysRevD.107.084005}{{\em
  Phys. Rev. D} {\bfseries 107} no.~8, (2023) 084005},
  \href{http://arxiv.org/abs/2301.06325}{{\ttfamily arXiv:2301.06325 [gr-qc]}}.

\bibitem{Calzetta:1993qe}
E.~Calzetta and B.~L. Hu, ``{Noise and fluctuations in semiclassical
  gravity},'' \href{http://dx.doi.org/10.1103/PhysRevD.49.6636}{{\em Phys. Rev.
  D} {\bfseries 49} (1994) 6636--6655},
  \href{http://arxiv.org/abs/gr-qc/9312036}{{\ttfamily arXiv:gr-qc/9312036}}.

\bibitem{Hu:1999mm}
B.~L. Hu, ``{Stochastic gravity},''
  \href{http://dx.doi.org/10.1023/A:1026664317157}{{\em Int. J. Theor. Phys.}
  {\bfseries 38} (1999) 2987--3037},
  \href{http://arxiv.org/abs/gr-qc/9902064}{{\ttfamily arXiv:gr-qc/9902064}}.

\bibitem{Hu:2008rga}
B.~L. Hu and E.~Verdaguer, ``{Stochastic Gravity: Theory and Applications},''
  \href{http://dx.doi.org/10.12942/lrr-2008-3}{{\em Living Rev. Rel.}
  {\bfseries 11} (2008) 3}, \href{http://arxiv.org/abs/0802.0658}{{\ttfamily
  arXiv:0802.0658 [gr-qc]}}.

\bibitem{Hertzberg:2021rbl}
M.~P. Hertzberg and J.~A. Litterer, ``{Bound on quantum fluctuations in
  gravitational waves from LIGO-Virgo},''
  \href{http://dx.doi.org/10.1088/1475-7516/2023/03/009}{{\em JCAP} {\bfseries
  03} (2023) 009}, \href{http://arxiv.org/abs/2112.12159}{{\ttfamily
  arXiv:2112.12159 [gr-qc]}}.

\bibitem{Kanno:2021gpt}
S.~Kanno, J.~Soda, and J.~Tokuda, ``{Indirect detection of gravitons through
  quantum entanglement},''
  \href{http://dx.doi.org/10.1103/PhysRevD.104.083516}{{\em Phys. Rev. D}
  {\bfseries 104} no.~8, (2021) 083516},
  \href{http://arxiv.org/abs/2103.17053}{{\ttfamily arXiv:2103.17053 [gr-qc]}}.

\bibitem{Arani:2021oim}
F.~S. Arani, M.~B. Harouni, B.~Lamine, and A.~Blanchard, ``{Sensing quantum
  nature of primordial gravitational waves using electromagnetic probes},''
  \href{http://dx.doi.org/10.1088/1402-4896/acb7ac}{{\em Phys. Scripta}
  {\bfseries 98} no.~5, (2023) 055004},
  \href{http://arxiv.org/abs/2110.10962}{{\ttfamily arXiv:2110.10962 [gr-qc]}}.

\bibitem{Giovannini:2010xg}
M.~Giovannini, ``{Hanbury Brown-Twiss interferometry and second-order
  correlations of inflaton quanta},''
  \href{http://dx.doi.org/10.1103/PhysRevD.83.023515}{{\em Phys. Rev. D}
  {\bfseries 83} (2011) 023515},
  \href{http://arxiv.org/abs/1011.1673}{{\ttfamily arXiv:1011.1673
  [astro-ph.CO]}}.

\bibitem{Kanno:2018cuk}
S.~Kanno and J.~Soda, ``{Detecting nonclassical primordial gravitational waves
  with Hanbury-Brown\textendash{}Twiss interferometry},''
  \href{http://dx.doi.org/10.1103/PhysRevD.99.084010}{{\em Phys. Rev. D}
  {\bfseries 99} no.~8, (2019) 084010},
  \href{http://arxiv.org/abs/1810.07604}{{\ttfamily arXiv:1810.07604
  [hep-th]}}.

\bibitem{Kanno:2019gqw}
S.~Kanno, ``{Nonclassical primordial gravitational waves from the initial
  entangled state},'' \href{http://dx.doi.org/10.1103/PhysRevD.100.123536}{{\em
  Phys. Rev. D} {\bfseries 100} no.~12, (2019) 123536},
  \href{http://arxiv.org/abs/1905.06800}{{\ttfamily arXiv:1905.06800
  [hep-th]}}.

\bibitem{Guerreiro:2019vbq}
T.~Guerreiro, ``{Quantum Effects in Gravity Waves},''
  \href{http://dx.doi.org/10.1088/1361-6382/ab9d5d}{{\em Class. Quant. Grav.}
  {\bfseries 37} no.~15, (2020) 155001},
  \href{http://arxiv.org/abs/1911.11593}{{\ttfamily arXiv:1911.11593
  [quant-ph]}}.

\bibitem{Guerreiro:2021qgk}
T.~Guerreiro, F.~Coradeschi, A.~M. Frassino, J.~R. West, and E.~Schioppa,
  Junior., ``{Quantum signatures in nonlinear gravitational waves},''
  \href{http://dx.doi.org/10.22331/q-2022-12-19-879}{{\em Quantum} {\bfseries
  6} (2022) 879}, \href{http://arxiv.org/abs/2111.01779}{{\ttfamily
  arXiv:2111.01779 [gr-qc]}}.

\bibitem{Hu:1992ig}
B.~L. Hu, J.~P. Paz, and Y.~Zhang, ``{Quantum origin of noise and fluctuations
  in cosmology},'' in {\em {The Origin of Structure in the Universe}}.
\newblock 1992.
\newblock \href{http://arxiv.org/abs/gr-qc/9512049}{{\ttfamily
  arXiv:gr-qc/9512049}}.

\bibitem{Calzetta:1995ys}
E.~Calzetta and B.~L. Hu, ``{Quantum fluctuations, decoherence of the mean
  field, and structure formation in the early universe},''
  \href{http://dx.doi.org/10.1103/PhysRevD.52.6770}{{\em Phys. Rev. D}
  {\bfseries 52} (1995) 6770--6788},
  \href{http://arxiv.org/abs/gr-qc/9505046}{{\ttfamily arXiv:gr-qc/9505046}}.

\bibitem{Lombardo:2005iz}
F.~C. Lombardo and D.~Lopez~Nacir, ``{Decoherence during inflation: The
  Generation of classical inhomogeneities},''
  \href{http://dx.doi.org/10.1103/PhysRevD.72.063506}{{\em Phys. Rev. D}
  {\bfseries 72} (2005) 063506},
  \href{http://arxiv.org/abs/gr-qc/0506051}{{\ttfamily arXiv:gr-qc/0506051}}.

\bibitem{Kiefer:2006je}
C.~Kiefer, I.~Lohmar, D.~Polarski, and A.~A. Starobinsky, ``{Pointer states for
  primordial fluctuations in inflationary cosmology},''
  \href{http://dx.doi.org/10.1088/0264-9381/24/7/002}{{\em Class. Quant. Grav.}
  {\bfseries 24} (2007) 1699--1718},
  \href{http://arxiv.org/abs/astro-ph/0610700}{{\ttfamily
  arXiv:astro-ph/0610700}}.

\bibitem{Kiefer:2008ku}
C.~Kiefer and D.~Polarski, ``{Why do cosmological perturbations look classical
  to us?},'' \href{http://dx.doi.org/10.1166/asl.2009.1023}{{\em Adv. Sci.
  Lett.} {\bfseries 2} (2009) 164--173},
  \href{http://arxiv.org/abs/0810.0087}{{\ttfamily arXiv:0810.0087
  [astro-ph]}}.

\bibitem{Burgess:2006jn}
C.~P. Burgess, R.~Holman, and D.~Hoover, ``{Decoherence of inflationary
  primordial fluctuations},''
  \href{http://dx.doi.org/10.1103/PhysRevD.77.063534}{{\em Phys. Rev. D}
  {\bfseries 77} (2008) 063534},
  \href{http://arxiv.org/abs/astro-ph/0601646}{{\ttfamily
  arXiv:astro-ph/0601646}}.

\bibitem{Martineau:2006ki}
P.~Martineau, ``{On the decoherence of primordial fluctuations during
  inflation},'' \href{http://dx.doi.org/10.1088/0264-9381/24/23/006}{{\em
  Class. Quant. Grav.} {\bfseries 24} (2007) 5817--5834},
  \href{http://arxiv.org/abs/astro-ph/0601134}{{\ttfamily
  arXiv:astro-ph/0601134}}.

\bibitem{Prokopec:2006fc}
T.~Prokopec and G.~I. Rigopoulos, ``{Decoherence from Isocurvature
  perturbations in Inflation},''
  \href{http://dx.doi.org/10.1088/1475-7516/2007/11/029}{{\em JCAP} {\bfseries
  11} (2007) 029}, \href{http://arxiv.org/abs/astro-ph/0612067}{{\ttfamily
  arXiv:astro-ph/0612067}}.

\bibitem{Burgess:2014eoa}
C.~P. Burgess, R.~Holman, G.~Tasinato, and M.~Williams, ``{EFT Beyond the
  Horizon: Stochastic Inflation and How Primordial Quantum Fluctuations Go
  Classical},'' \href{http://dx.doi.org/10.1007/JHEP03(2015)090}{{\em JHEP}
  {\bfseries 03} (2015) 090}, \href{http://arxiv.org/abs/1408.5002}{{\ttfamily
  arXiv:1408.5002 [hep-th]}}.

\bibitem{Boyanovsky:2015tba}
D.~Boyanovsky, ``{Effective field theory during inflation: Reduced density
  matrix and its quantum master equation},''
  \href{http://dx.doi.org/10.1103/PhysRevD.92.023527}{{\em Phys. Rev. D}
  {\bfseries 92} no.~2, (2015) 023527},
  \href{http://arxiv.org/abs/1506.07395}{{\ttfamily arXiv:1506.07395
  [astro-ph.CO]}}.

\bibitem{Nelson:2016kjm}
E.~Nelson, ``{Quantum Decoherence During Inflation from Gravitational
  Nonlinearities},''
  \href{http://dx.doi.org/10.1088/1475-7516/2016/03/022}{{\em JCAP} {\bfseries
  03} (2016) 022}, \href{http://arxiv.org/abs/1601.03734}{{\ttfamily
  arXiv:1601.03734 [gr-qc]}}.

\bibitem{Shandera:2017qkg}
S.~Shandera, N.~Agarwal, and A.~Kamal, ``{Open quantum cosmological system},''
  \href{http://dx.doi.org/10.1103/PhysRevD.98.083535}{{\em Phys. Rev. D}
  {\bfseries 98} no.~8, (2018) 083535},
  \href{http://arxiv.org/abs/1708.00493}{{\ttfamily arXiv:1708.00493
  [hep-th]}}.

\bibitem{Martin:2018zbe}
J.~Martin and V.~Vennin, ``{Observational constraints on quantum decoherence
  during inflation},''
  \href{http://dx.doi.org/10.1088/1475-7516/2018/05/063}{{\em JCAP} {\bfseries
  05} (2018) 063}, \href{http://arxiv.org/abs/1801.09949}{{\ttfamily
  arXiv:1801.09949 [astro-ph.CO]}}.

\bibitem{Martin:2018lin}
J.~Martin and V.~Vennin, ``{Non Gaussianities from Quantum Decoherence during
  Inflation},'' \href{http://dx.doi.org/10.1088/1475-7516/2018/06/037}{{\em
  JCAP} {\bfseries 06} (2018) 037},
  \href{http://arxiv.org/abs/1805.05609}{{\ttfamily arXiv:1805.05609
  [astro-ph.CO]}}.

\bibitem{DaddiHammou:2022itk}
A.~Daddi~Hammou and N.~Bartolo, ``{Cosmic decoherence: primordial power spectra
  and non-Gaussianities},'' \href{http://arxiv.org/abs/2211.07598}{{\ttfamily
  arXiv:2211.07598 [astro-ph.CO]}}.

\bibitem{Colas:2022hlq}
T.~Colas, J.~Grain, and V.~Vennin, ``{Benchmarking the cosmological master
  equations},'' \href{http://dx.doi.org/10.1140/epjc/s10052-022-11047-9}{{\em
  Eur. Phys. J. C} {\bfseries 82} no.~12, (2022) 1085},
  \href{http://arxiv.org/abs/2209.01929}{{\ttfamily arXiv:2209.01929
  [hep-th]}}.

\bibitem{Ye:2018kty}
G.~Ye and Y.-S. Piao, ``{Quantum decoherence of primordial perturbations
  through nonlinear scaler-tensor interaction},''
  \href{http://arxiv.org/abs/1806.07672}{{\ttfamily arXiv:1806.07672 [gr-qc]}}.

\bibitem{Sou:2022nsd}
C.~M. Sou, D.~H. Tran, and Y.~Wang, ``{Decoherence of cosmological
  perturbations from boundary terms and the non-classicality of gravity},''
  \href{http://dx.doi.org/10.1007/JHEP04(2023)092}{{\em JHEP} {\bfseries 04}
  (2023) 092}, \href{http://arxiv.org/abs/2207.04435}{{\ttfamily
  arXiv:2207.04435 [hep-th]}}.

\bibitem{Friedrich:2019hev}
P.~Friedrich and T.~Prokopec, ``{Entropy production in inflation from spectator
  loops},'' \href{http://dx.doi.org/10.1103/PhysRevD.100.083505}{{\em Phys.
  Rev. D} {\bfseries 100} no.~8, (2019) 083505},
  \href{http://arxiv.org/abs/1907.13564}{{\ttfamily arXiv:1907.13564
  [astro-ph.CO]}}.

\bibitem{Brahma:2020zpk}
S.~Brahma, O.~Alaryani, and R.~Brandenberger, ``{Entanglement entropy of
  cosmological perturbations},''
  \href{http://dx.doi.org/10.1103/PhysRevD.102.043529}{{\em Phys. Rev. D}
  {\bfseries 102} no.~4, (2020) 043529},
  \href{http://arxiv.org/abs/2005.09688}{{\ttfamily arXiv:2005.09688
  [hep-th]}}.

\bibitem{Gong:2019yyz}
J.-O. Gong and M.-S. Seo, ``{Quantum non-linear evolution of inflationary
  tensor perturbations},''
  \href{http://dx.doi.org/10.1007/JHEP05(2019)021}{{\em JHEP} {\bfseries 05}
  (2019) 021}, \href{http://arxiv.org/abs/1903.12295}{{\ttfamily
  arXiv:1903.12295 [hep-th]}}.

\bibitem{Burgess:2022nwu}
C.~P. Burgess, R.~Holman, G.~Kaplanek, J.~Martin, and V.~Vennin, ``{Minimal
  decoherence from inflation},''
  \href{http://arxiv.org/abs/2211.11046}{{\ttfamily arXiv:2211.11046
  [hep-th]}}.

\bibitem{Kanno:2022ykw}
S.~Kanno, J.~Soda, and K.~Ueda, ``{Conversion of squeezed gravitons into
  photons during inflation},''
  \href{http://dx.doi.org/10.1103/PhysRevD.106.083508}{{\em Phys. Rev. D}
  {\bfseries 106} no.~8, (2022) 083508},
  \href{http://arxiv.org/abs/2207.05734}{{\ttfamily arXiv:2207.05734
  [hep-th]}}.

\bibitem{Kanno:2022kve}
S.~Kanno, J.~Soda, and K.~Ueda, ``{Impact of quantum entanglement induced by
  magnetic fields on primordial gravitational waves},''
  \href{http://arxiv.org/abs/2211.05576}{{\ttfamily arXiv:2211.05576
  [hep-th]}}.

\bibitem{Arroja:2011yj}
F.~Arroja and T.~Tanaka, ``{A note on the role of the boundary terms for the
  non-Gaussianity in general k-inflation},''
  \href{http://dx.doi.org/10.1088/1475-7516/2011/05/005}{{\em JCAP} {\bfseries
  05} (2011) 005}, \href{http://arxiv.org/abs/1103.1102}{{\ttfamily
  arXiv:1103.1102 [astro-ph.CO]}}.

\bibitem{Burrage:2011hd}
C.~Burrage, R.~H. Ribeiro, and D.~Seery, ``{Large slow-roll corrections to the
  bispectrum of noncanonical inflation},''
  \href{http://dx.doi.org/10.1088/1475-7516/2011/07/032}{{\em JCAP} {\bfseries
  07} (2011) 032}, \href{http://arxiv.org/abs/1103.4126}{{\ttfamily
  arXiv:1103.4126 [astro-ph.CO]}}.

\bibitem{Prokopec:2013zya}
T.~Prokopec and J.~Weenink, ``{Frame independent cosmological perturbations},''
  \href{http://dx.doi.org/10.1088/1475-7516/2013/09/027}{{\em JCAP} {\bfseries
  09} (2013) 027}, \href{http://arxiv.org/abs/1304.6737}{{\ttfamily
  arXiv:1304.6737 [gr-qc]}}.

\bibitem{DeWitt:1967yk}
B.~S. DeWitt, ``{Quantum Theory of Gravity. 1. The Canonical Theory},''
  \href{http://dx.doi.org/10.1103/PhysRev.160.1113}{{\em Phys. Rev.} {\bfseries
  160} (1967) 1113--1148}.

\bibitem{Wheeler:1968iap}
J.~A. Wheeler, ``{SUPERSPACE AND THE NATURE OF QUANTUM GEOMETRODYNAMICS},''
  {\em Adv. Ser. Astrophys. Cosmol.} {\bfseries 3} (1987) 27--92.

\bibitem{Witten:2022xxp}
E.~Witten, ``{A Note On The Canonical Formalism for Gravity},''
  \href{http://arxiv.org/abs/2212.08270}{{\ttfamily arXiv:2212.08270
  [hep-th]}}.

\bibitem{Araujo-Regado:2022gvw}
G.~Araujo-Regado, R.~Khan, and A.~C. Wall, ``{Cauchy slice holography: a new
  AdS/CFT dictionary},'' \href{http://dx.doi.org/10.1007/JHEP03(2023)026}{{\em
  JHEP} {\bfseries 03} (2023) 026},
  \href{http://arxiv.org/abs/2204.00591}{{\ttfamily arXiv:2204.00591
  [hep-th]}}.

\bibitem{McGough:2016lol}
L.~McGough, M.~Mezei, and H.~Verlinde, ``{Moving the CFT into the bulk with $
  T\overline{T} $},'' \href{http://dx.doi.org/10.1007/JHEP04(2018)010}{{\em
  JHEP} {\bfseries 04} (2018) 010},
  \href{http://arxiv.org/abs/1611.03470}{{\ttfamily arXiv:1611.03470
  [hep-th]}}.

\bibitem{Hartman:2018tkw}
T.~Hartman, J.~Kruthoff, E.~Shaghoulian, and A.~Tajdini, ``{Holography at
  finite cutoff with a $T^2$ deformation},''
  \href{http://dx.doi.org/10.1007/JHEP03(2019)004}{{\em JHEP} {\bfseries 03}
  (2019) 004}, \href{http://arxiv.org/abs/1807.11401}{{\ttfamily
  arXiv:1807.11401 [hep-th]}}.

\bibitem{Araujo-Regado:2022jpj}
G.~Araujo-Regado, ``{Holographic Cosmology on Closed Slices in 2+1
  Dimensions},'' \href{http://arxiv.org/abs/2212.03219}{{\ttfamily
  arXiv:2212.03219 [hep-th]}}.

\bibitem{Larsen:2004kf}
F.~Larsen and R.~McNees, ``{Holography, diffeomorphisms, and scaling violations
  in the CMB},'' \href{http://dx.doi.org/10.1088/1126-6708/2004/07/062}{{\em
  JHEP} {\bfseries 07} (2004) 062},
  \href{http://arxiv.org/abs/hep-th/0402050}{{\ttfamily arXiv:hep-th/0402050}}.

\bibitem{Pimentel:2013gza}
G.~L. Pimentel, ``{Inflationary Consistency Conditions from a Wavefunctional
  Perspective},'' \href{http://dx.doi.org/10.1007/JHEP02(2014)124}{{\em JHEP}
  {\bfseries 02} (2014) 124}, \href{http://arxiv.org/abs/1309.1793}{{\ttfamily
  arXiv:1309.1793 [hep-th]}}.

\bibitem{Pajer:2016ieg}
E.~Pajer, G.~L. Pimentel, and J.~V.~S. Van~Wijck, ``{The Conformal Limit of
  Inflation in the Era of CMB Polarimetry},''
  \href{http://dx.doi.org/10.1088/1475-7516/2017/06/009}{{\em JCAP} {\bfseries
  06} (2017) 009}, \href{http://arxiv.org/abs/1609.06993}{{\ttfamily
  arXiv:1609.06993 [hep-th]}}.

\bibitem{Larsen:2003pf}
F.~Larsen and R.~McNees, ``{Inflation and de Sitter holography},''
  \href{http://dx.doi.org/10.1088/1126-6708/2003/07/051}{{\em JHEP} {\bfseries
  07} (2003) 051}, \href{http://arxiv.org/abs/hep-th/0307026}{{\ttfamily
  arXiv:hep-th/0307026}}.

\bibitem{Maldacena:2019cbz}
J.~Maldacena, G.~J. Turiaci, and Z.~Yang, ``{Two dimensional Nearly de Sitter
  gravity},'' \href{http://dx.doi.org/10.1007/JHEP01(2021)139}{{\em JHEP}
  {\bfseries 01} (2021) 139}, \href{http://arxiv.org/abs/1904.01911}{{\ttfamily
  arXiv:1904.01911 [hep-th]}}.

\bibitem{Chakraborty:2023yed}
T.~Chakraborty, J.~Chakravarty, V.~Godet, P.~Paul, and S.~Raju, ``{The Hilbert
  space of de Sitter quantum gravity},''
  \href{http://arxiv.org/abs/2303.16315}{{\ttfamily arXiv:2303.16315
  [hep-th]}}.

\bibitem{Chakraborty:2023los}
T.~Chakraborty, J.~Chakravarty, V.~Godet, P.~Paul, and S.~Raju, ``{Holography
  of information in de Sitter space},''
  \href{http://arxiv.org/abs/2303.16316}{{\ttfamily arXiv:2303.16316
  [hep-th]}}.

\bibitem{Freidel:2008sh}
L.~Freidel, ``{Reconstructing AdS/CFT},''
  \href{http://arxiv.org/abs/0804.0632}{{\ttfamily arXiv:0804.0632 [hep-th]}}.

\bibitem{Papadimitriou:2010as}
I.~Papadimitriou, ``{Holographic renormalization as a canonical
  transformation},'' \href{http://dx.doi.org/10.1007/JHEP11(2010)014}{{\em
  JHEP} {\bfseries 11} (2010) 014},
  \href{http://arxiv.org/abs/1007.4592}{{\ttfamily arXiv:1007.4592 [hep-th]}}.

\bibitem{Maldacena:2002vr}
J.~M. Maldacena, ``{Non-Gaussian features of primordial fluctuations in single
  field inflationary models},''
  \href{http://dx.doi.org/10.1088/1126-6708/2003/05/013}{{\em JHEP} {\bfseries
  05} (2003) 013}, \href{http://arxiv.org/abs/astro-ph/0210603}{{\ttfamily
  arXiv:astro-ph/0210603}}.

\bibitem{York:1972sj}
J.~W. York, Jr., ``{Role of conformal three geometry in the dynamics of
  gravitation},'' \href{http://dx.doi.org/10.1103/PhysRevLett.28.1082}{{\em
  Phys. Rev. Lett.} {\bfseries 28} (1972) 1082--1085}.

\bibitem{Gibbons:1976ue}
G.~W. Gibbons and S.~W. Hawking, ``{Action Integrals and Partition Functions in
  Quantum Gravity},'' \href{http://dx.doi.org/10.1103/PhysRevD.15.2752}{{\em
  Phys. Rev. D} {\bfseries 15} (1977) 2752--2756}.

\bibitem{York:1986lje}
J.~York, ``{Boundary terms in the action principles of general relativity},''
  \href{http://dx.doi.org/10.1007/BF01889475}{{\em Found. Phys.} {\bfseries 16}
  (1986) 249--257}.

\bibitem{Prokopec:2010be}
T.~Prokopec and G.~Rigopoulos, ``{Path Integral for Inflationary
  Perturbations},'' \href{http://dx.doi.org/10.1103/PhysRevD.82.023529}{{\em
  Phys. Rev. D} {\bfseries 82} (2010) 023529},
  \href{http://arxiv.org/abs/1004.0882}{{\ttfamily arXiv:1004.0882 [gr-qc]}}.

\bibitem{Ota:2022xni}
A.~Ota, M.~Sasaki, and Y.~Wang, ``{One-loop tensor power spectrum from an
  excited scalar field during inflation},''
  \href{http://arxiv.org/abs/2211.12766}{{\ttfamily arXiv:2211.12766
  [astro-ph.CO]}}.

\bibitem{Maldacena:2011nz}
J.~M. Maldacena and G.~L. Pimentel, ``{On graviton non-Gaussianities during
  inflation},'' \href{http://dx.doi.org/10.1007/JHEP09(2011)045}{{\em JHEP}
  {\bfseries 09} (2011) 045}, \href{http://arxiv.org/abs/1104.2846}{{\ttfamily
  arXiv:1104.2846 [hep-th]}}.

\bibitem{Rigopoulos:2011eq}
G.~Rigopoulos, ``{Gauge invariance and non-Gaussianity in Inflation},''
  \href{http://dx.doi.org/10.1103/PhysRevD.84.021301}{{\em Phys. Rev. D}
  {\bfseries 84} (2011) 021301},
  \href{http://arxiv.org/abs/1104.0292}{{\ttfamily arXiv:1104.0292
  [astro-ph.CO]}}.

\bibitem{Prokopec:2012ug}
T.~Prokopec and J.~Weenink, ``{Uniqueness of the gauge invariant action for
  cosmological perturbations},''
  \href{http://dx.doi.org/10.1088/1475-7516/2012/12/031}{{\em JCAP} {\bfseries
  12} (2012) 031}, \href{http://arxiv.org/abs/1209.1701}{{\ttfamily
  arXiv:1209.1701 [gr-qc]}}.

\bibitem{Garcia-Saenz:2019njm}
S.~Garcia-Saenz, L.~Pinol, and S.~Renaux-Petel, ``{Revisiting non-Gaussianity
  in multifield inflation with curved field space},''
  \href{http://dx.doi.org/10.1007/JHEP01(2020)073}{{\em JHEP} {\bfseries 01}
  (2020) 073}, \href{http://arxiv.org/abs/1907.10403}{{\ttfamily
  arXiv:1907.10403 [hep-th]}}.

\bibitem{Wang:2013mea}
Y.~Wang, ``{MathGR: a tensor and GR computation package to keep it simple},''
  \href{http://arxiv.org/abs/1306.1295}{{\ttfamily arXiv:1306.1295 [cs.MS]}}.

\bibitem{Wang:2013zva}
Y.~Wang, ``{Inflation, Cosmic Perturbations and Non-Gaussianities},''
  \href{http://dx.doi.org/10.1088/0253-6102/62/1/19}{{\em Commun. Theor. Phys.}
  {\bfseries 62} (2014) 109--166},
  \href{http://arxiv.org/abs/1303.1523}{{\ttfamily arXiv:1303.1523 [hep-th]}}.

\bibitem{Salopek:1992qy}
D.~S. Salopek and J.~M. Stewart, ``{Hamilton-Jacobi theory for general
  relativity with matter fields},''
  \href{http://dx.doi.org/10.1088/0264-9381/9/8/015}{{\em Class. Quant. Grav.}
  {\bfseries 9} (1992) 1943--1968}.

\bibitem{Parry:1993mw}
J.~Parry, D.~S. Salopek, and J.~M. Stewart, ``{Solving the Hamilton-Jacobi
  equation for general relativity},''
  \href{http://dx.doi.org/10.1103/PhysRevD.49.2872}{{\em Phys. Rev. D}
  {\bfseries 49} (1994) 2872--2881},
  \href{http://arxiv.org/abs/gr-qc/9310020}{{\ttfamily arXiv:gr-qc/9310020}}.

\bibitem{Liu:2016aaf}
J.~Liu, C.-M. Sou, and Y.~Wang, ``{Cosmic Decoherence: Massive Fields},''
  \href{http://dx.doi.org/10.1007/JHEP10(2016)072}{{\em JHEP} {\bfseries 10}
  (2016) 072}, \href{http://arxiv.org/abs/1608.07909}{{\ttfamily
  arXiv:1608.07909 [hep-th]}}.

\bibitem{Barvinsky:1998cq}
A.~O. Barvinsky, A.~Y. Kamenshchik, C.~Kiefer, and I.~V. Mishakov,
  ``{Decoherence in quantum cosmology at the onset of inflation},''
  \href{http://dx.doi.org/10.1016/S0550-3213(99)00208-4}{{\em Nucl. Phys. B}
  {\bfseries 551} (1999) 374--396},
  \href{http://arxiv.org/abs/gr-qc/9812043}{{\ttfamily arXiv:gr-qc/9812043}}.

\bibitem{Planck:2018jri}
{\bfseries Planck} Collaboration, Y.~Akrami {\em et~al.}, ``{Planck 2018
  results. X. Constraints on inflation},''
  \href{http://dx.doi.org/10.1051/0004-6361/201833887}{{\em Astron. Astrophys.}
  {\bfseries 641} (2020) A10},
  \href{http://arxiv.org/abs/1807.06211}{{\ttfamily arXiv:1807.06211
  [astro-ph.CO]}}.

\bibitem{tHooft:1974toh}
G.~'t~Hooft and M.~J.~G. Veltman, ``{One loop divergencies in the theory of
  gravitation},'' {\em Ann. Inst. H. Poincare Phys. Theor. A} {\bfseries 20}
  (1974) 69--94.

\bibitem{Witten:2018lgb}
E.~Witten, ``{A note on boundary conditions in Euclidean gravity},''
  \href{http://dx.doi.org/10.1142/S0129055X21400043}{{\em Rev. Math. Phys.}
  {\bfseries 33} no.~10, (2021) 2140004},
  \href{http://arxiv.org/abs/1805.11559}{{\ttfamily arXiv:1805.11559
  [hep-th]}}.

\bibitem{FRADKIN1984472}
E.~Fradkin and A.~Tseytlin, ``One-loop effective potential in gauged o(4)
  supergravity and the problem of the $\lambda$ term,''
  \href{http://dx.doi.org/https://doi.org/10.1016/0550-3213(84)90074-9}{{\em
  Nuclear Physics B} {\bfseries 234} no.~2, (1984) 472--508}.

\bibitem{Bamba:2014mua}
K.~Bamba, G.~Cognola, S.~D. Odintsov, and S.~Zerbini, ``{One-loop modified
  gravity in a de Sitter universe, quantum-corrected inflation, and its
  confrontation with the Planck result},''
  \href{http://dx.doi.org/10.1103/PhysRevD.90.023525}{{\em Phys. Rev. D}
  {\bfseries 90} no.~2, (2014) 023525},
  \href{http://arxiv.org/abs/1404.4311}{{\ttfamily arXiv:1404.4311 [gr-qc]}}.

\bibitem{Barvinsky:2006gd}
A.~O. Barvinsky, ``{Quantum Effective Action in Spacetimes with Branes and
  Boundaries: Diffeomorphism Invariance},''
  \href{http://dx.doi.org/10.1103/PhysRevD.74.084033}{{\em Phys. Rev. D}
  {\bfseries 74} (2006) 084033},
  \href{http://arxiv.org/abs/hep-th/0608004}{{\ttfamily arXiv:hep-th/0608004}}.

\bibitem{Gibbons:1978ac}
G.~W. Gibbons, S.~W. Hawking, and M.~J. Perry, ``{Path Integrals and the
  Indefiniteness of the Gravitational Action},''
  \href{http://dx.doi.org/10.1016/0550-3213(78)90161-X}{{\em Nucl. Phys. B}
  {\bfseries 138} (1978) 141--150}.

\bibitem{Christensen:1979iy}
S.~M. Christensen and M.~J. Duff, ``{Quantizing Gravity with a Cosmological
  Constant},'' \href{http://dx.doi.org/10.1016/0550-3213(80)90423-X}{{\em Nucl.
  Phys. B} {\bfseries 170} (1980) 480--506}.

\bibitem{Weinberg:2005vy}
S.~Weinberg, ``{Quantum contributions to cosmological correlations},''
  \href{http://dx.doi.org/10.1103/PhysRevD.72.043514}{{\em Phys. Rev. D}
  {\bfseries 72} (2005) 043514},
  \href{http://arxiv.org/abs/hep-th/0506236}{{\ttfamily arXiv:hep-th/0506236}}.

\bibitem{Senatore:2009cf}
L.~Senatore and M.~Zaldarriaga, ``{On Loops in Inflation},''
  \href{http://dx.doi.org/10.1007/JHEP12(2010)008}{{\em JHEP} {\bfseries 12}
  (2010) 008}, \href{http://arxiv.org/abs/0912.2734}{{\ttfamily arXiv:0912.2734
  [hep-th]}}.

\bibitem{Pimentel:2012tw}
G.~L. Pimentel, L.~Senatore, and M.~Zaldarriaga, ``{On Loops in Inflation III:
  Time Independence of zeta in Single Clock Inflation},''
  \href{http://dx.doi.org/10.1007/JHEP07(2012)166}{{\em JHEP} {\bfseries 07}
  (2012) 166}, \href{http://arxiv.org/abs/1203.6651}{{\ttfamily arXiv:1203.6651
  [hep-th]}}.

\bibitem{Senatore:2012nq}
L.~Senatore and M.~Zaldarriaga, ``{On Loops in Inflation II: IR Effects in
  Single Clock Inflation},''
  \href{http://dx.doi.org/10.1007/JHEP01(2013)109}{{\em JHEP} {\bfseries 01}
  (2013) 109}, \href{http://arxiv.org/abs/1203.6354}{{\ttfamily arXiv:1203.6354
  [hep-th]}}.

\bibitem{Brahma:2022yxu}
S.~Brahma, A.~Berera, and J.~Calder\'on-Figueroa, ``{Quantum corrections to the
  primordial tensor spectrum: open EFTs \& Markovian decoupling of UV modes},''
  \href{http://dx.doi.org/10.1007/JHEP08(2022)225}{{\em JHEP} {\bfseries 08}
  (2022) 225}, \href{http://arxiv.org/abs/2206.05797}{{\ttfamily
  arXiv:2206.05797 [hep-th]}}.

\bibitem{DelCampo:2019afl}
A.~Del~Campo and T.~Takayanagi, ``{Decoherence in Conformal Field Theory},''
  \href{http://dx.doi.org/10.1007/JHEP02(2020)170}{{\em JHEP} {\bfseries 02}
  (2020) 170}, \href{http://arxiv.org/abs/1911.07861}{{\ttfamily
  arXiv:1911.07861 [hep-th]}}.

\bibitem{Cicoli:2023opf}
M.~Cicoli, J.~P. Conlon, A.~Maharana, S.~Parameswaran, F.~Quevedo, and
  I.~Zavala, ``{String Cosmology: from the Early Universe to Today},''
  \href{http://arxiv.org/abs/2303.04819}{{\ttfamily arXiv:2303.04819
  [hep-th]}}.

\end{thebibliography}\endgroup

\end{document}